\documentclass[prd,showpacs,superscriptaddress,twocolumn,floatfix,nofootinbib,preprintnumbers,altaffilletter, letterpaper]{revtex4-1}

\usepackage{amsmath,amssymb}
\usepackage{xspace}
\usepackage{xcolor}
\usepackage{graphicx}
\usepackage{multirow}
\usepackage{enumitem}
\graphicspath{{.} {Figures/}}
\usepackage{color}
\usepackage{amssymb}
\usepackage{amsmath}
\usepackage{times}
\usepackage{bm}
\usepackage{array}
\usepackage{acronym}
\usepackage[para,online,flushleft]{threeparttable}
\usepackage{colortbl}
\usepackage{booktabs}
\usepackage[colorlinks=true,citecolor=teal]{hyperref}
\usepackage[T1]{fontenc}

\newcommand{\orcid}[1]{\href{https://orcid.org/#1}{\includegraphics[width=10pt]{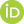}}}

\newcommand{\mlstat}{\mbox{MLStat}\xspace}
\newcommand{\pycbc}{{\sc PyCBC}\xspace}
\newcommand{\gstlal}{{\sc GstLAL}\xspace}
\newcommand{\mbta}{{\sc MBTA}\xspace}
\newcommand{\cwb}{{\sc cWB}\xspace}
\newcommand{\bilby}{{\sc bilby}\xspace}
\newcommand{\bayeswave}{{\sc BayesWave}\xspace}
\newcommand{\waveform}{{\sc SEOBNRv4HM\_ROM}\xspace}
\newcommand{\Pcbc}{{$\mathrm{P}_\mathrm{CBC}$}\xspace}

\newcommand{\msun}{\text{M}_{\odot}\xspace}
\newcommand{\mpc}{\text{Mpc}\xspace}

\newcommand{\newevent}{GW151216\xspace}

\newcommand{\massone}{46.8^{+12.5}_{-17.7}\msun\xspace}
\newcommand{\masstwo}{23.6^{+5.6}_{-9.6}\msun\xspace}
\newcommand{\chieff}{\chi_\mathrm{eff}\xspace}
\newcommand{\chieffval}{0.56^{+0.23}_{-0.5}\xspace}
\newcommand{\lumdist}{2269^{+1193}_{-1033}\mpc\xspace}

\begin{document}

\title{Improving significance of binary black hole mergers in Advanced LIGO data using deep learning : Confirmation of \newevent}

\author{Shreejit Jadhav\orcid{0000-0003-0554-0084}}
\email{shreejit@iucaa.in}
\affiliation{Inter-University Centre for Astronomy and Astrophysics (IUCAA), Post Bag 4, Ganeshkhind, Pune 411 007, India}
\author{Nikhil Mukund\orcid{0000-0002-8666-9156}} 
\email{nikhil.mukund@aei.mpg.de}
\affiliation{Max-Planck-Institut f{\"u}r Gravitationsphysik (Albert-Einstein-Institut) and Institut f{\"u}r Gravitationsphysik, \\ Leibniz Universit{\"a}t Hannover, Callinstra{\ss}e 38, 30167 Hannover, Germany }
\affiliation{Inter-University Centre for Astronomy and Astrophysics (IUCAA), Post Bag 4, Ganeshkhind, Pune 411 007, India}
\author{Bhooshan Gadre\orcid{0000-0002-1534-9761}}
\email{bhooshan.gadre@aei.mpg.de}
\affiliation{Max Planck Institute for Gravitational Physics (Albert Einstein Institute), Am M\"uhlenberg 1, D-14476 Potsdam-Golm, Germany}
\affiliation{Inter-University Centre for Astronomy and Astrophysics (IUCAA), Post Bag 4, Ganeshkhind, Pune 411 007, India}
\author{Sanjit Mitra\orcid{0000-0002-0800-4626}}
\email{sanjit@iucaa.in}
\affiliation{Inter-University Centre for Astronomy and Astrophysics (IUCAA), Post Bag 4, Ganeshkhind, Pune 411 007, India}
\author{Sheelu Abraham\orcid{0000-0001-9524-2739}}
\email{sheelu@iucaa.in}
\affiliation{Inter-University Centre for Astronomy and Astrophysics (IUCAA), Post Bag 4, Ganeshkhind, Pune 411 007, India}
\date{\today}

\begin{abstract}
We present a novel Machine Learning (ML) based strategy to search for binary black hole (BBH) mergers
in data from ground-based gravitational wave (GW) observatories.
This is the first ML-based search that not only recovers all the compact binary coalescences (CBCs) in the first GW transients catalog (GWTC-1), but also makes a clean detection of \newevent, which was not significant enough to be included in the catalog. Moreover, we achieve this by only adding a new coincident ranking statistic (\mlstat) to a standard analysis that was used for \mbox{GWTC-1}. In CBC searches, reducing contamination by terrestrial and instrumental transients, which create a loud noise background by triggering numerous false alarms, is crucial to improving the sensitivity for detecting true events. The sheer volume of data and a large number of expected detections also prompts the use of ML techniques. We perform transfer learning to train ``InceptionV3'', a pre-trained deep neural network, along with curriculum learning to distinguish GW signals from noisy events by analysing their continuous wavelet transform (CWT) maps. \mlstat incorporates information from this ML classifier into the coincident search likelihood used by the standard \pycbc search. This leads to at least an order of magnitude improvement in the inverse false-alarm-rate (IFAR) for the previously ``low significance'' events GW151012, GW170729 and \newevent. We also perform the parameter estimation of \newevent using \waveform. We carry out an injection study to show that \mlstat brings substantial improvement to the detection sensitivity of Advanced LIGO for all compact binary coalescences. The average improvement in the sensitive volume is $\sim 10\%$ for low chirp masses ($0.8-5~\msun$), and $\sim 30\%$ for higher masses ($5-50~\msun$). Performance in the lower masses may become even better if the training set for the ML classifier, currently restricted to black hole binaries with component masses in the range $2-98~\msun$ only, is expanded to include binaries with Neutron stars. Considering the impressive ability of the statistic to distinguish signals from glitches, the list of marginal events from \mlstat could be quite reliable for astrophysical population studies and further follow-up. This work demonstrates the immense potential and readiness of \mlstat for finding new sources in current data and the possibility of its adaptation in similar searches.
\end{abstract}

\pacs{}
\maketitle

\section{Introduction}

The Advanced LIGO~\cite{aLIGO, aLIGO_Detectors} and the Advanced Virgo~\cite{aVIRGO} observatories have reported the detection of fifty compact binary coalescences (CBCs) so far~\cite{GW150914, GW151226, GW170104, GW170608, GW170814, gwtc1, gwtc2}. They consist of binary black-hole (BBH) mergers including one with asymmetric component masses~\cite{GW190412}, another one resulting in the production of an intermediate mass black hole (IMBH)~\cite{GW190521}, a possible neutron star - black hole (NSBH) merger~\cite{GW190814} and a couple of binary neutron star mergers~\cite{GW170817, GW190425} in their data spanning the first two observational runs (O1, O2) and the first half of the third observational run (O3a). As the LIGO and Virgo detectors advance their sensitivities and with KAGRA joining the search~\cite{Kagradesign, akutsu2020overview}, the rate of GW detection is expected to increase multifold~\cite{abbott2018prospects}.

Currently, these detectors make use of both modelled and unmodelled searches to detect potential GW signals in their calibrated data~\cite{Usman_2016, sachdev2019gstlal, klimenko2016method, adams2015low, 2020arXiv201106787C}. Had the noise in the detectors been Gaussian and stationary, only the time-coincidence of GW signals in more than one detector would be a decisive criterion to find a candidate event. The quadrature sum of signal-to-noise ratios (SNRs) in individual detectors would suffice as the ranking statistic~\cite{wainsteinextraction, Cutler, Pai, Allen12} and modelling the background analytically would be possible in this simplistic case~\cite{Finn, Finn_Chern}. However, the data is heavily contaminated by non-stationary and non-Gaussian transients of instrumental and terrestrial origin, also known as `glitches'. These noise transients mimic the astrophysical signals and result in a persistent problem of false alarms~\cite{abbott2016characterization, nuttall2015improving} which impact the search sensitivity of coalescing compact binaries and GW bursts. As the detectors become more sensitive, along with an improvement in detection rate of GW signals the rate of occurrence of such glitches may also increase, resulting in a considerable number of real GW candidates being reported with reduced significance.

Various methods of compact binary coalescence search~\cite{Usman_2016, messick2017analysis, klimenko2016method} were used to detect gravitational wave sources in LIGO's first observational run (O1)~\cite{abbott2016binary, abbott2017search}. These searches resulted in two confident BBH detections viz., GW150914 and GW151226, each at a false-alarm rate (FAR) of $< 6.0 \times 10^{-7}\ \mathrm{yr}^{-1}$ ($> 5.3 \sigma$), and a much less confident third detection viz. GW151012. The \pycbc analysis detected GW151012 with a FAR of $0.37\ \mathrm{yr}^{-1}$ ($1.7 \sigma$)~\cite{GW150914, abbott2016gw150914, BBH_O1}. Later, using an improved ranking statistic, this value could be reduced to $0.17\ \mathrm{yr}^{-1}$~\cite{gwtc1}. Similarly, in the offline \pycbc analysis of the second observational run (O2), GW170729 was detected with a FAR of $1.36\ \mathrm{yr}^{-1}$.\footnote{Note that \gstlal detected GW151012 and GW170729 with better significance than \pycbc. In \cwb, GW151012 could not be detected, while GW170729 was detected with a significance higher than \pycbc and \gstlal~\cite{BBH_O1, gwtc1}.} The statistic used in 2-OGC~\cite{2OGC} reduced the FAR values of GW151012 and GW170729 to $0.0045$ and $0.15$ respectively. An investigation shows that the triggers with SNR $\approx 9$ mainly consist of false alarms caused by glitches~\cite{BBH_O1}, which increase the noise background and reduce the significance of marginal events, thereby obstructing the science we can do with them.
 
In this work we demonstrate how application of a machine learning algorithm, designed to discern real events from spurious glitches, can be used to improve the standard matched filtering based analysis used by LIGO~\cite{gwtc1}. We use transfer learning with InceptionV3~\cite{szegedy2016rethinking} to classify CBCs and glitches in the LIGO data. We construct a new coincident ranking statistic (\textit{\mlstat}) by incorporating the output of the ML classifier into the original ranking statistic. Here we extend the \pycbc search analysis used in GWTC-1~\cite{improved_pycbc, gwtc1}, however, a similar methodology can also be implemented with other pipelines like \cwb~\cite{klimenko2016method}, \gstlal~\cite{sachdev2019gstlal} and \mbta~\cite{adams2015low} to seek sensitivity improvements. We show significant improvement in the inverse-false-alarm rates (IFARs) of the low-significance events GW151012 and GW170729 with \mlstat. We also confirm the event \newevent from O1~\cite{IAS_O1, IAS_GW151216} and report the parameter estimation results for completeness.

In the past few years, machine learning methods have been applied in GW data analysis extensively~\cite{Matching_MF, BNS_ML, cuoco2020enhancing, PE_AutoRegrFl, DNN_MMA, 2018PhLB..778...64G, Detecting_SN_DL, SimilarityLearning, 10.1093/mnras/staa3550, PhysRevD.101.083006, 2019PhRvD.100f2005M, 2019PhRvD.100f3015G}. Though, most of the works have relied on simulated data, there have been efforts in recent years to implement ML methods on real LIGO data~\cite{10.1093/mnras/staa3550, PhysRevD.101.083006, 2019PhRvD.100f2005M, 2018PhLB..778...64G}. However, careful considerations are required while working with real data. For example, the results discussed in~\cite{2018PhLB..778...64G} appear convincing but the data considered in the work are only limited to $4096$ seconds around the three events from O1 and thus, may not prove equally effective on longer durations of real LIGO data which contain a much wider variety of noise artefacts. Also, as discussed in~\cite{2019PhRvD.100f3015G}, applying deep learning models directly on a continuous stream of whitened data may require additional post-processing. The performance metrics of the ML models should be evaluated using a data handling scheme that is a priori blind to the locations of the injections made. As our analysis is based on triggers obtained after matched-filtering, which localises the time of merger for true events, this issue is avoided. We are not aware of any ML based search that has been demonstrated to be able to catch all the GWTC-1 events.

The paper is organised as follows. Section~\ref{sec:incorp_ml} introduces the state-of-the-art deep learning model, viz., InceptionV3, describes its implementation on the LIGO data using transfer learning and gives details about training and validation. Section~\ref{sec:construct_mlstat} describes the construction and operation of \mlstat. Results of the injection study used to test \mlstat are given in Subsection~\ref{sec:injstudy}. In Section~\ref{sec:analysis}, we give details of the analysis of data from first and second observing runs. The Subsection~\ref{sec:ext_analysis} discusses the approximation method introduced by us to gauge the significance improvement of the full foreground of first and second observing runs. Section~\ref{sec:conclusion} concludes the paper.

\section{Incorporating Machine Learning}\label{sec:incorp_ml}

Success for any ML classifier depends on its ability to learn and optimally separate the relevant features from the dataset. Within astronomy community, features are traditionally chosen by a domain expert as in the case of several stellar~\cite{Manteiga_2009, S_nchez_Almeida_2013, kuntzer}, gamma-ray burst~\cite{abraham2020machine}, galaxy~\cite{star_galaxy_dbnn, auto_galax_morph, randforest}, quasar~\cite{Richards_2008, abraham_sdss, Peters_2015} and GW classification schemes~\cite{dbnn, gravityspy}. The advent of graphics processing units (GPUs) and the availability of larger training sets have resulted in techniques based on deep learning to gain more prominence in recent years. Convolutional neural networks (CNNs), in particular, have shown remarkable success in tasks pertaining to image classification and are consistently outperforming feature-based classical ML techniques when applied to standard bench-marking datasets~\cite{deng2009imagenet,krizhevsky2012imagenet}. Large architectures of CNNs, like Google's Inception model and Microsoft's Resnet model, have been trained using cutting-edge GPU technology on large datasets containing images of objects in day-to-day life~\cite{szegedy2016rethinking, resnet}. Transfer learning makes use of the rich feature extraction capability of such pre-trained models and allows re-purposing them for a different classification task. The final few layers of a pre-trained network are replaced with trainable new layers while the rest of the network is retained as before. The training of such a network on a new image dataset essentially maps the extracted features to new classes, thus making the training faster while still maintaining the accuracy. Another advantage of transfer learning is that the amount of training data required to reach the prescribed accuracy is hugely reduced.

Inception networks~\cite{szegedy2015going} put forward a strategy for making effective deeper networks through tactics different from merely increasing either the layers or neurons per layer. For example, the InceptionV3 network~\cite{szegedy2016rethinking} differs from the traditional monolithic CNN architecture through the usage of factorising convolutions, parallel structures, and extensive dimensionality reduction techniques. These schemes make the architecture well suited for applications with stringent memory and computational constraints but still provide state-of-the-art performance without over-fitting the data.

The transients of astrophysical or terrestrial origin seen in LIGO data have a peculiar evolution in the time-frequency domain. This provides a 2D representation that can be analysed by the image-based classifiers mentioned above. During the early inspiral phase of a CBC waveform, the evolution of frequency with time can be approximated as \mbox{$f \propto (t_c - t) ^{-3/8}$}, where $t_c$ is the time of coalescence~\cite{creighton2012gravitational}. This explains the peculiar chirp-like shape of CBC signals in time-frequency maps which can be exploited to differentiate them from glitches. We get much more accurate description of the full time-frequency evolution of CBCs using the Inspiral-Merger-Ringdown (IMR) waveforms. We use transfer learning along with curriculum learning to retrain InceptionV3 for the classification of continuous wavelet transform (CWT) maps of transients in the LIGO data. For each evaluated image, the network outputs a posterior probability across the $17$ transient classes used for training. However, only the probability corresponding to CBC class (\Pcbc) will be used in our analysis. The probability corresponding to other glitch classes may also be incorporated in the future analyses provided the leakage between CBC and other glitch classes is carefully evaluated.

Previously, Omega scans have been used to map the LIGO data to the time-frequency domain where the chirp-like evolution of CBCs can be distinguished from other noise transients more effectively~\cite{gravityspy,dbnn}. We use the continuous wavelet transform (CWT) with an analytic Morlet (amor) wavelet to construct time-frequency scalograms of the whitened strain data to be analysed. Wavelets, in general, provide a much better time and frequency resolution compared to short-time Fourier transform (STFT)~\cite{debnath2002wavelet}. Discriminator based on relative wavelet energy has previously been demonstrated to be effective in separating various transient classes and was successfully applied to Advanced LIGO's first observational run data~\cite{dbnn}. We create CWT scalograms by whitening and band-passing the data around a GPS trigger between $16$~Hz and $512$~Hz. We consider a data slice of $1$ second duration with the trigger time kept at centre, convert it into a scalogram and save it as a grayscale image with pixels denoting absolute values of CWT coefficients. As InceptionV3 is trained on natural images, the features extracted from different channels are most likely to differ from each other based on the biases in the images of natural objects. The choice of using grayscale colormap ensures the complete glitch morphology is saved in each of the three channels rather than getting divided based on the colormap. The network's convolution filters thus observe the full evolution of a transient in each channel and the channel-based biases are marginalised.

The CWT image data were generated by using a subset of GravitySpy data~\cite{gravityspy, GravitySpyDb}. The data were manually curated to remove images that lacked clear features or where the features could not be covered in the $1$ second long CWT scalograms. The data were then split into training ($70\%$) and validation ($30\%$) sets. As transfer learning involves training a significantly fewer number of neurons, typically, a very small amount of training data (up to a thousand images per class) suffices. This size also makes manual curation practically possible. Due to band passing, classes like 1080Lines, 1400Ripples and Violin-Mode were rendered redundant and were thus excluded. Also, the Power-Line class was divided into Power-Line and Power-Line2 to separate shorter and longer transients. Data for CBC class was generated by considering real LIGO strain data and by injecting CBC signals having chirp-distances\footnote{chirp-distance accounts for the leading order chirp mass dependence in the amplitude of a CBC signal; see Appendix F in~\cite{PhysRevD.79.122001}} between $5$ Mpc and $300$ Mpc with SEOBNRv4 waveform approximant~\cite{seobnrv4}. The component masses were sampled between $(2\ \mathrm{M}_\odot, 98\ \mathrm{M}_\odot)$ with a constraint on total mass, $M \leq 100\ \mathrm{M}_\odot$. Though some of the low mass binaries would merge at frequencies higher than $512$~Hz, a major part of these signals contributing most of the power would still be covered by the allowed frequency band. Besides, the upper cutoff of $512$~Hz eliminates the relatively louder noise transients at higher frequencies and allows the classifier to focus on the part of signal lying in the most sensitive frequency band of LIGO. A separate class named Gaussian-Noise was created, which consisted of plain whitened strain data that did not contain any transients. The classes Koi Fish, No Glitch, Paired Doves and None of the Above were omitted from the data due to similarity with at least one of the other classes or lack of feature distinction or coverage in the $1$ second long CWT maps. A total of $17$ transient classes have been used, viz., Air Compressor (AC), Blip (BL), Extremely Loud (EL), Gaussian Noise (GN), Helix (HX), Injected Chirp (CBC), Light Modulation (LM), Low Frequency Burst (LFB), Low Frequency Lines (LFL), Power Line (PL1), Power Line2 (PL2), Repeating Blips (RBL), Scattered Light (SL), Scratchy (SC), Tomte (TM), Wandering Line (WL) and Whistle (WH).

\begin{figure}
    \includegraphics[trim=0.5cm 0cm 1cm 1cm, clip=true, width=\linewidth]{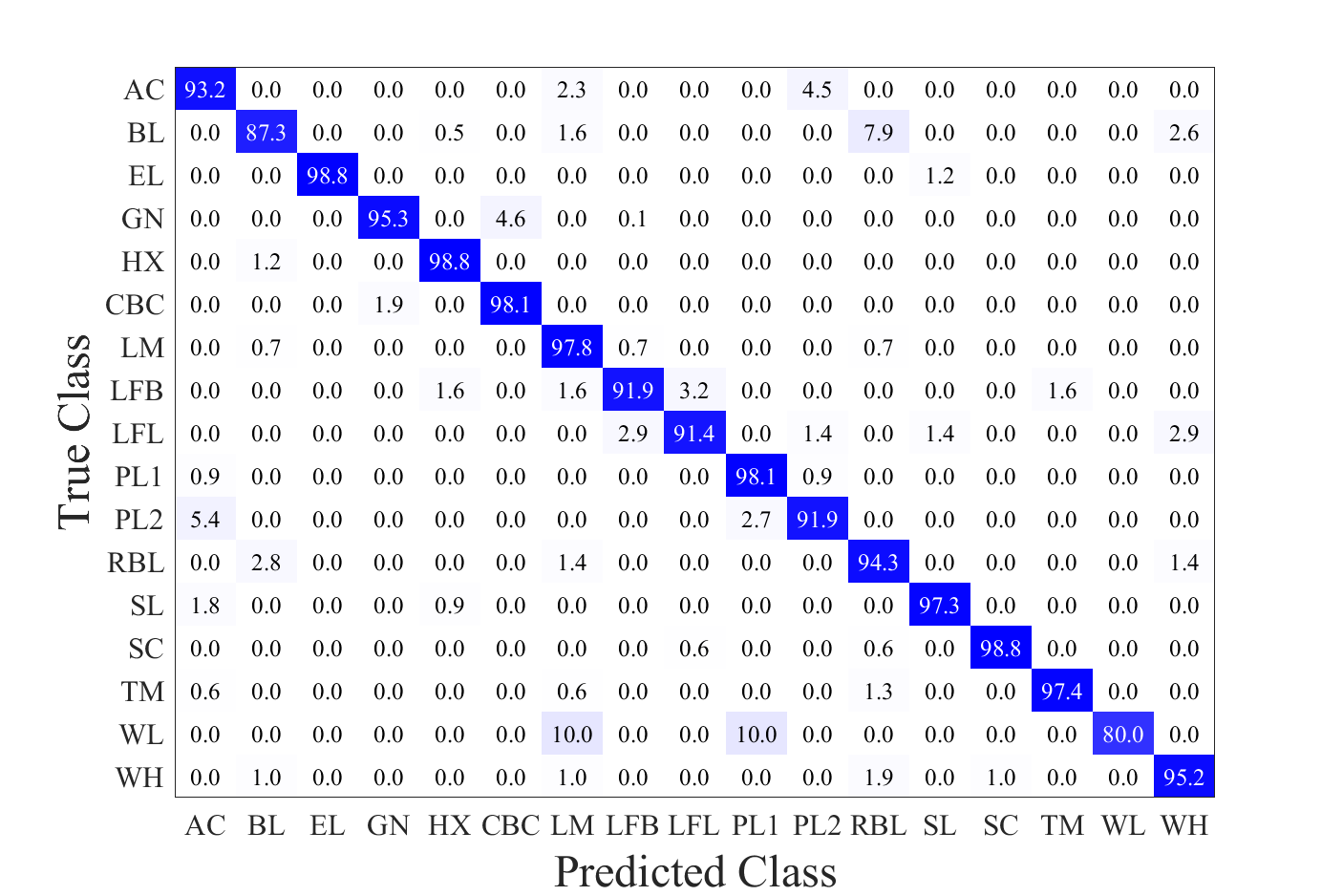}
    \caption{Confusion matrix (in percent) for the combined validation data from all the levels of curriculum learning for $17$ transient classes. The performance of CBC class is of particular importance to \mlstat.
    }
    \label{fig:conf_mat}
\end{figure}

\begin{figure}
    \includegraphics[trim=0cm 0cm 1cm 1cm, clip=true, width=\linewidth]{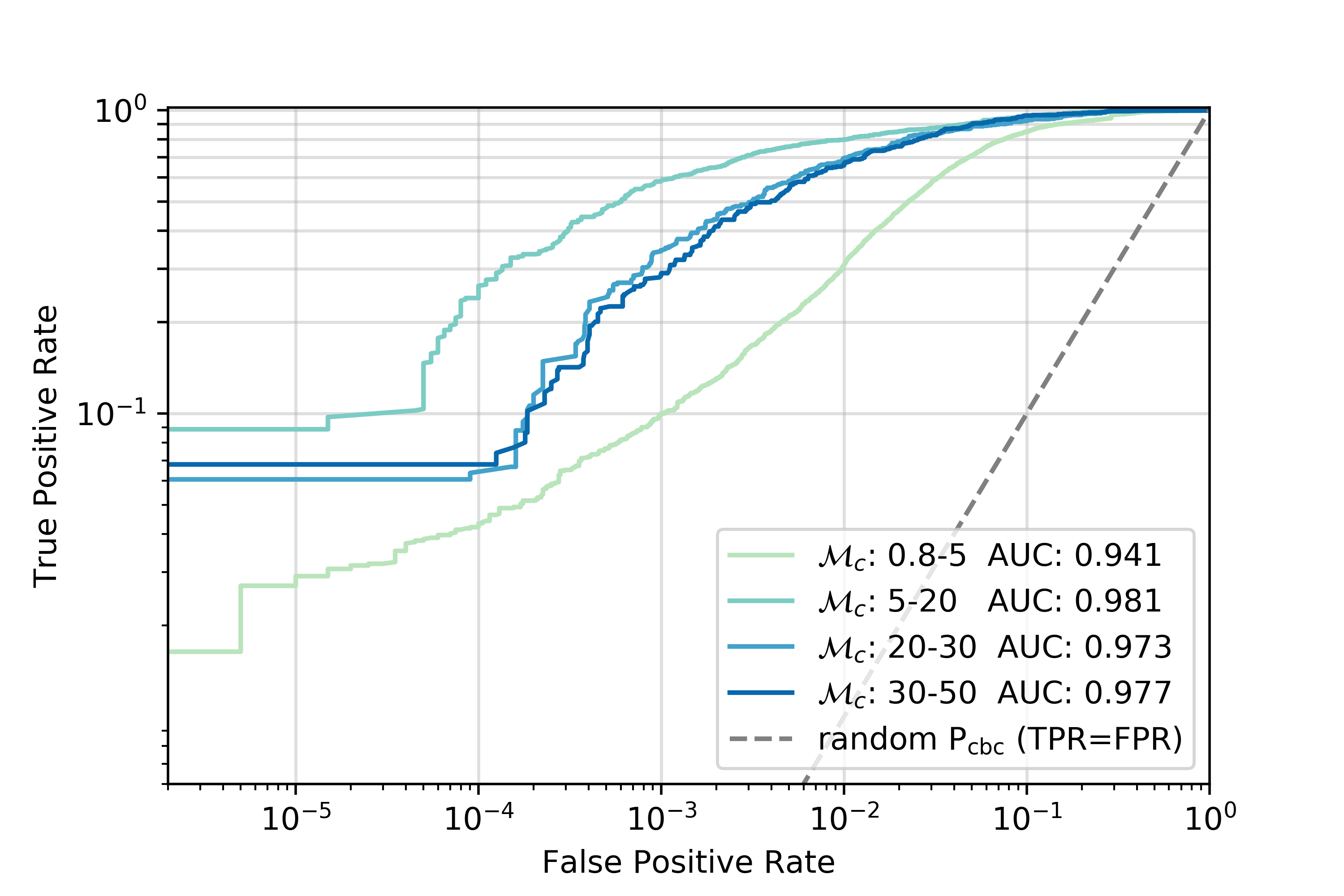}
    \caption{The receiver operating characteristic curves for different chirp mass bins obtained by comparing simulated injections against randomly chosen noise triggers from Hanford and Livingston data. Log scales are used to visualise the differences better. Masses for the noise triggers are taken from the respective templates triggering them. As some injections lie out of the training range, the performance in the $0.8-5~\msun$ chirp mass bin is suboptimal as compared to other bins.
    }
    \label{fig:roc}
\end{figure}

To perform transfer learning, we load the 316 layers deep directed acyclic graph (DAG) network of \textit{InceptionV3} and freeze the weights of the first 250 layers. We also add a $40\%$ \textit{Dropout} layer and an additional \textit{Fully Connected} layer of size $1024$ with \textit{Leaky-ReLu} activation function before the final replaced \textit{Fully Connected} layer which now maps to the the $17$ transient classes with \textit{Softmax} activation function. We adopt a curriculum learning strategy specifically in CBC class to train the network step-by-step, going from higher SNR values to lower ones. We observed that this method remarkably increased the network accuracy as compared to a single training session with full data. The network was trained using stochastic gradient descent with momentum (SGD-M)~\cite{qian1999momentum}. After three levels of training, the trained network achieved a training accuracy of $100\%$ and a validation accuracy of $95.8\%$ 
(note that the data being curated manually, very high accuracies are expected). In the last level of learning, the training and validation accuracies plateau after the $12^\mathrm{th}$ epoch and show only statistical fluctuations. We analysed all the checkpoints of the network recorded at the end of every epoch and chose the best checkpoint based on overall validation accuracy. In Fig.~\ref{fig:conf_mat}, we show the confusion matrix for the entire validation data. A high percentage of predictions lie on the diagonal indicating satisfactory performance of the classifier. Fig.~\ref{fig:roc} shows the receiver operating characteristic (ROC) curve for our network, demonstrating its efficiency of detecting injections against other triggers (mostly originating from noise). Though we do not use the ROC curve to set a threshold on \Pcbc, the high \textit{area\ under\ the\ curve} (AUC) scores are indicative of the network's ability to distinguish signals across all chirp mass bins from glitches.

\section{Construction of \mlstat}\label{sec:construct_mlstat}

We build our ML tool as an augmentation of one of the standard pipelines used by the LIGO-VIRGO Collaboration, viz. \pycbc. The \pycbc workflow performs matched filtering on the data with a bank of templates. Triggers are then collected by thresholding and clustering the SNR time series. For each trigger, the SNR is re-weighted with two types of noise suppressing vetoes~\cite{Allen_chi, sgveto} and a single detector ranking statistic is calculated ensuring approximately constant trigger rate across the search parameter space~\cite{improved_pycbc}. The pipeline then finds triggers coincident in Hanford and Livingston detectors with an allowed time window of $15$ ms, which accounts for the light travel time and uncertainty in recorded times of coalescence. A semi-coherent ranking statistic (we call it base statistic hereafter) is evaluated for these foreground triggers~\cite{improved_pycbc}. To construct the background, the same procedure is repeated between the two detectors with time shifted triggers. A false-alarm-rate which decides the significance of each foreground trigger is then evaluated based on this background. 

We analyse the triggers collected by the \pycbc workflow with our classifier. For each trigger, our classifier gives the probability (\Pcbc) of it belonging to the CBC class. The likelihood ratio for the detection of a GW signal in the given data from the $i^\mathrm{th}$ detector is,

\begin{equation} \label{eq:old_liklihood}
    \Lambda(\mathcal{H}_\lambda|s_i) = \frac{p(s_i|\mathcal{H}_\lambda)}{p(s_i|\mathcal{H}_0)}, 
\end{equation}

where, $s_i(t)$ is the time series strain data, $\mathcal{H}_\lambda$ is the hypothesis stating that a signal $h_i(t, \lambda)$ is present in $s_i(t)$ and $\mathcal{H}_0$ is the null hypothesis. We update the above likelihood ratio to include \Pcbc as follows\footnote{The factor $(\mathrm{P}_\mathrm{NOISE})_i$ was not included in the denominator of the likelihood ratio to maintain the constraint $\tilde{\varrho}_\mathrm{ml} \leq \tilde{\rho}_\mathrm{c}$.},

\begin{equation} \label{eq:new_liklihood}
    \Lambda'(\mathcal{H}_\lambda|s_i) = \frac{p(s_i|\mathcal{H}_\lambda) * (\mathrm{P}_\mathrm{CBC})_\mathrm{i}}{p(s_i|\mathcal{H}_0)}.
\end{equation}

Thus the combined likelihood ratio manifests a new coincident ranking statistic (\mlstat) $\tilde{\varrho}_\mathrm{ml}$, a simple extension to the coincident base statistic $\tilde{\rho}_\mathrm{c}$, as,

\begin{equation} \label{eq:newstat}
    \tilde{\varrho}^2_\mathrm{ml} = \tilde{\rho}^2_\mathrm{c} + 2 * \mathrm{log}[(\mathrm{P}_\mathrm{CBC})_\mathrm{c}],
\end{equation}

where, the combined probability,

\begin{equation}
    (\mathrm{P}_\mathrm{CBC})_\mathrm{c} = (\mathrm{P}_\mathrm{CBC})_\mathrm{H} * (\mathrm{P}_\mathrm{CBC})_\mathrm{L}.
\end{equation}

Here, H and L in the subscripts denote Hanford and Livingston observatories respectively. Thus, for a detection with \mlstat, the signal should look CBC-like in both the detectors within the time window required for coincidence\footnote{Note that we do not use \Pcbc for fitting the single detector event rate as done in the base statistic which may give us further improvements, especially in the regions of the parameter space which can suffer due to high false-alarm rates and thus show reduced sensitivity.}.

The astrophysical CBC signals should have \Pcbc values close to $1$, thus making the magnitude of second term on the right hand side of Eq. \ref{eq:newstat} very small. Thus, the \mlstat values will be very close to the base statistic for such events. On the other hand, noise triggers in the background which do not have a CBC-like composition in CWT maps will be pushed to lower values. Thus, discriminating real events from the noise contamination results in a decreased background, which effectively improves the significance of true CBC events. However, for louder signals, even the low \Pcbc values in both the detectors will not result in any significant difference between the base statistic and \mlstat. This works in our favour while recovering events like GW170817 (shown in Table~\ref{tab:All_events}) which lie outside our present training range of the network but are recovered with negligible reduction in the statistic. The ultimate target of \mlstat would be recovering the marginal events in the base search.

\subsection{Injection Study}\label{sec:injstudy}

\begin{figure}[htb]
    \includegraphics[clip=true, width=\linewidth]{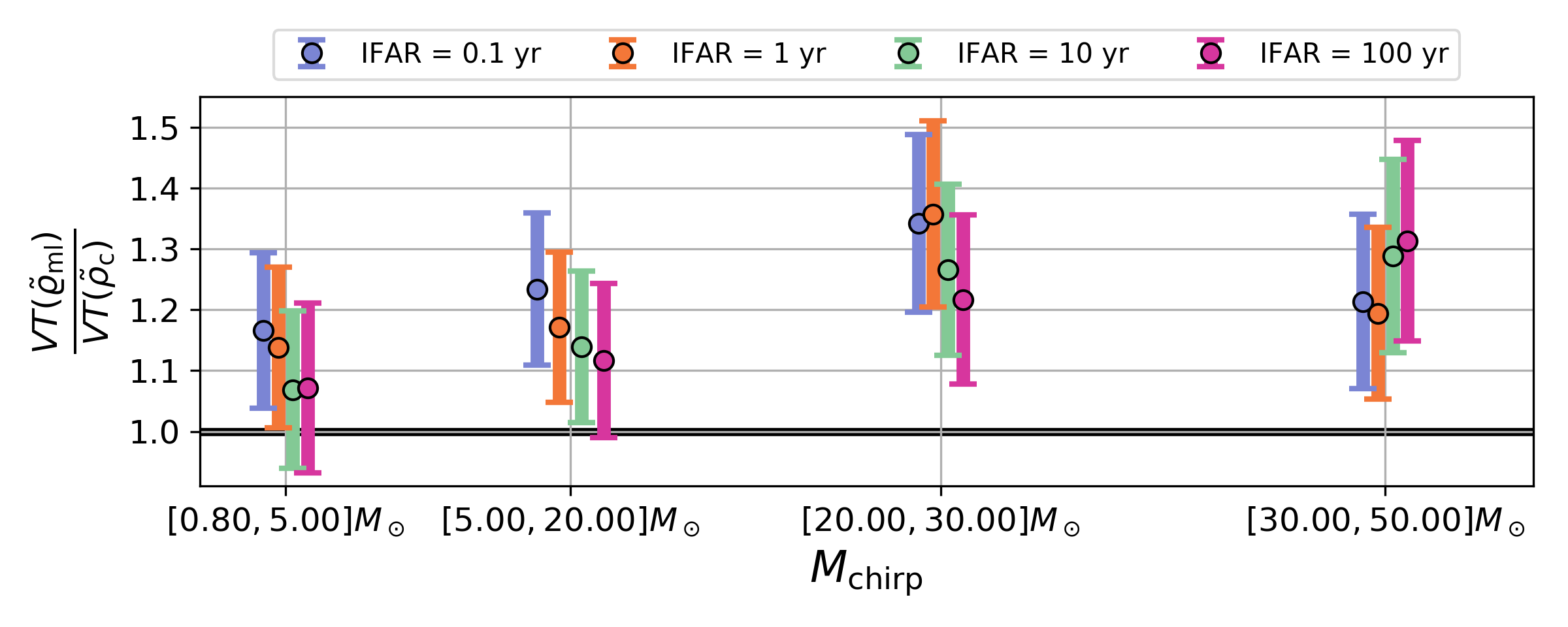}
    \includegraphics[clip=true, trim=0cm 0cm 0cm 1.2cm, width=\linewidth]{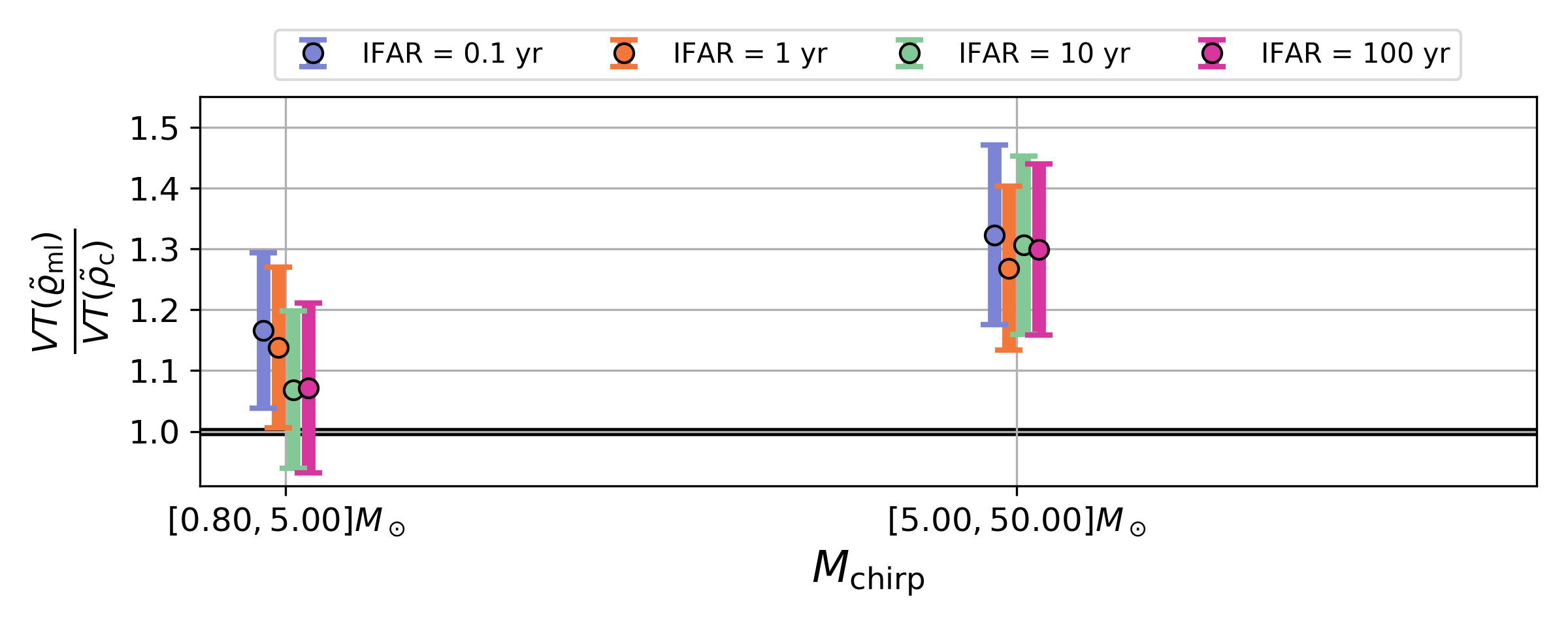}
    \caption{Ratio of the sensitive volume-time (VT) estimates with \mlstat and base statistic for different IFAR thresholds and four chirp mass bins obtained from the injection study is shown in the top panel. The bottom panel combines the higher mass bins to reduce the errorbars, showing an average $\sim 10\%$ improvement in sensitive volume for low chirp masses ($< 5\ \msun$) and overall $\sim 30\%$ improvement for higher chirp masses.
    }
    \label{fig:vt}
\end{figure}

To test the performance of \mlstat rigorously, an injection study was performed in three chunks of LIGO data containing the events GW151012, GW151216 and GW170729 respectively. In total, $5720$ simulated BBH and BNS merger signals, isotropic in sky locations and inclinations, having chirp distances between $5$ Mpc to $300$ Mpc, were injected in the real LIGO strain. The masses for BBH injections were sampled such that the distribution of total mass was uniform up to $100\ \mathrm{M}_\odot$ while component masses were between $2\ \mathrm{M}_\odot$ and $98\ \mathrm{M}_\odot$. Whereas, for BNS injections the component masses were sampled uniformly in the range $1\ \mathrm{M}_\odot$ to $3\ \mathrm{M}_\odot$. The waveform approximant used for BBH injections was SEOBNRv4~\cite{seobnrv4} while for BNS it was SpinTaylorT5~\cite{SpinTaylorT5}. The injections were recovered using search pipelines based on \mlstat and the ranking statistic used in \pycbc analysis of GWTC-1 (base statistic here). The IFAR values of the recovered injections were assigned using the background of the respective chunks and statistics. The sensitive spacetime volume $\langle \mathrm{VT} \rangle$ estimates were made in different chirp mass bins~\cite{gwtc1}. Ratios of the $\langle \mathrm{VT} \rangle$ estimates with \mlstat and base statistic are plotted in Fig.~\ref{fig:vt} for different chirp mass bins. An overall improvement of $~30\%$ in sensitivity can be observed for chirp masses above $5\ \mathrm{M}_\odot$. Though, for chirp masses below $5\ \mathrm{M}_\odot$, the improvement in sensitivity with \mlstat was not expected as the training as well as data pre-processing is limited to higher masses, the improvement is evident mainly because of the aggressive reduction in overall background. Thus, even if the low \Pcbc values result in lower values of \mlstat as compared to base statistic, the overall IFAR values for these injections show a small improvement.

\begin{table*}[ht]
    \caption{List of candidate events from the extended analysis of O1 and O2 sorted by FAR$_\mathrm{ml}$~\cite{gitlink}. We report one to two orders of magnitude improvements in IFAR for GW151012, GW170729 and \newevent, obtained by analysing the respective data chunks which makes them definite confident detections. The $\mathrm{FAR}_\mathrm{ml}$ values quoted for events lying outside these analysed chunks are inferred by assuming similar significance improvements across respective observational runs (see Section~\ref{sec:ext_analysis}). The table is split into two parts to show these events separately. All the catalog events~\cite{gwtc1} are successfully found by the combined analysis (exception is GW170818 as it was not found by \pycbc). GW170817 has very low \Pcbc values in both the detectors as it lies outside the training range of the network. Nevertheless, it is still recovered due to its loudness (details in Section~\ref{sec:construct_mlstat}). Three marginal events with $\mathrm{FAR}_\mathrm{ml}$ < $1$/month are included, which may be worth following-up and may prove to be useful for estimating astrophysical distributions. Mass and spin values (detector frame) are those of the best matching search template giving highest coincident \mlstat and need not match the full Bayesian parameter estimation results. $\mathrm{FAR}_\mathrm{base}$ values are also given for comparison. The newly detected event \newevent is shown in bold.
    }
    \centering
    \begin{threeparttable}
        \renewcommand\TPTminimum{3in} 
        \renewcommand{\arraystretch}{1}
        \begin{tabular}{%
        >{\raggedright\arraybackslash}p{0.16\linewidth}%
        >{\centering\arraybackslash}p{0.06\linewidth}%
        >{\raggedright\arraybackslash}p{0.12\linewidth}%
        >{\raggedright\arraybackslash}p{0.12\linewidth}%
        >{\raggedleft\arraybackslash}p{0.07\linewidth}%
        >{\raggedleft\arraybackslash}p{0.07\linewidth}%
        >{\raggedleft\arraybackslash}p{0.06\linewidth}%
        >{\raggedleft\arraybackslash}p{0.06\linewidth}%
        >{\raggedleft\arraybackslash}p{0.1\linewidth}%
        >{\raggedleft\arraybackslash}p{0.1\linewidth}}
            \hline\hline
            UTC & \mlstat & $\mathrm{FAR}_\mathrm{base}$ & $\mathrm{FAR}_\mathrm{ml}$ & $m_1^t$ & $m_2^t$ & $s_{2z}^t$ & $s_{1z}^t$ & $(\mathrm{P}_\mathrm{CBC})_\mathrm{H}$ & $(\mathrm{P}_\mathrm{CBC})_\mathrm{L}$ \\
            ~ & ~ & [yr$^{-1}$] & [yr$^{-1}$] & [M$_{\odot}$] & [M$_{\odot}$] & ~ & ~ & ~ & ~\\
            \hline
            2015-10-12T09:54:43\tnote{a, d} & 9.0 & 0.17 & $4.58 \times 10^{-3}$ & 25.75 & 17.60 & -0.36 & 0.73 & 0.889 & 0.996 \\
            2017-07-29T18:56:29\tnote{a, d} & 8.7 & 1.36 & $1.79 \times 10^{-2}$ & 67.52 & 32.53 & 0.27 & -0.09 & 0.836 & 1 \\
            {\bf 2015-12-16T09:24:16}\tnote{b, d} & {\bf 8.3} & {\bf 57.74} & {\bf 0.69} & {\bf 41.78} & {\bf 34.35} & {\bf 0.85} & {\bf 0.98} & {\bf 0.977} & {\bf 0.997} \\
            2017-07-28T04:54:39\tnote{d} & 7.9 & 410.49 & 3.08 & 33.1 & 2.50 & -0.85 & -0.45 & 0.793 & 0.990 \\
            2015-10-16T13:57:41\tnote{c, d} & 8.0 & ... & 4.80 & 364.0 & 5.35 & 0.98 & -0.24 & 0.998 & 0.885 \\
            \addlinespace[2ex]
            2017-08-17T12:41:04\tnote{a} & 28.3 & $<1.3 \times 10^{-5}$ & $<1.3 \times 10^{-5}$ & 1.46 & 1.30 & -0.02 & 0.01 & $1.7 \times 10^{-5}$ & $1.44 \times 10^{-3}$ \\
            2017-08-14T10:30:43\tnote{a} & 12.7 & $<1.3 \times 10^{-5}$ & $<1.3 \times 10^{-5}$ & 33.14 & 25.38 & 0.68 & -0.95 & 1 & 1 \\
            2017-01-04T10:11:58\tnote{a} & 10.8 & $<1.4 \times 10^{-5}$ & $<1.4 \times 10^{-5}$ & 40.87 & 13.91 & -0.70 & 0.80 & 0.999 & 1 \\
            2015-09-14T09:50:45\tnote{a} & 15.5 & $<1.5 \times 10^{-5}$ & $<1.5 \times 10^{-5}$ & 44.21 & 32.16 & 0.78 & -0.86 & 1 & 1 \\
            2015-12-26T03:38:53\tnote{a} & 11.8 & $<1.7 \times 10^{-5}$ & $<1.7 \times 10^{-5}$ & 14.83 & 8.50 & -0.09 & 0.81 & 0.849 & 1 \\
            2017-08-23T13:13:58\tnote{a} & 10.2 & $<3.3 \times 10^{-5}$ & $<3.3 \times 10^{-5}$ & 47.94 & 16.23 & -0.92 & 0.64 & 0.999 & 1 \\
            2017-08-09T08:28:21\tnote{a} & 10.5 & $<1.45 \times 10^{-4}$ & $<1.45 \times 10^{-4}$ & 47.62 & 16.21 & -0.57 & 0.91 & 0.983 & 1 \\
            2017-06-08T02:01:16\tnote{a} & 12.7 & $<3.1 \times 10^{-4}$ & $<3.1 \times 10^{-4}$ & 16.82 & 6.10 & 0.11 & 0.88 & 0.999 & 1 \\
            2016-12-14T16:26:40 & 8.0 & 847.26 & 6.02 & 15.09 & 11.77 & 0.79 & 0.54 & 0.996 & 0.973 \\
            \arrayrulecolor{black}
            \hline\hline
        \end{tabular}
        \begin{tablenotes}
            \raggedright
            \item [a] GWTC-1 events; \item [b] Not present in GWTC-1 but $\mathrm{FAR}_\mathrm{ml} < 1$ yr$^{-1}$; \item [c] Not present in the \pycbc analysis used in GWTC-1; \item [d] Definite $\mathrm{FAR}_\mathrm{ml}$ (obtained from analysed chunks)
        \end{tablenotes}
    \end{threeparttable}
    \label{tab:All_events}
\end{table*}

\section{Analysis of O1 and O2 data}\label{sec:analysis}

\begin{figure*}[htb]
    \centering
    \includegraphics[width=0.48\textwidth]{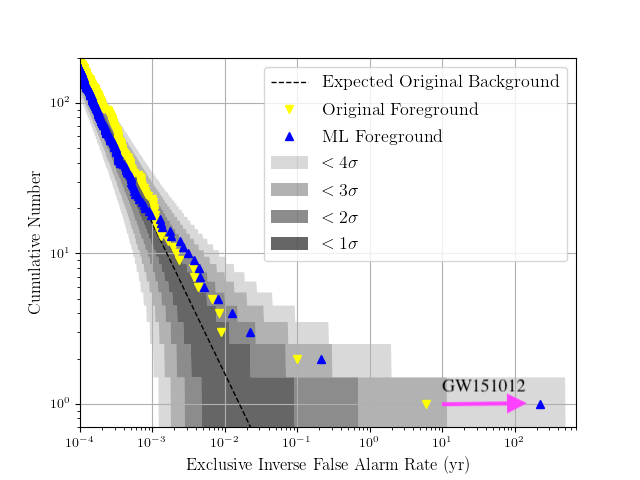}
    \includegraphics[width=0.48\textwidth]{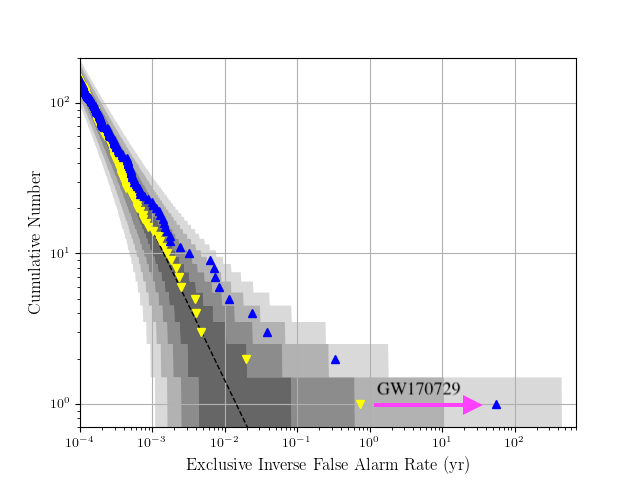}
    \caption{Significance improvement for GW151012 (left) and GW170729 (right): Cumulative histograms of foreground events in base statistic (yellow) and \mlstat (blue) along with the expected background plotted against the inverse false-alarm rate. Shaded regions show the sigma intervals for Poisson uncertainty. The loudness of first $\sim 20$ foreground events with \mlstat in O2 is not a systematic bias and is rather an effect of the low number statistics which has been observed with other statistics before (See Fig. 2 and Fig. 3 in~\cite{gwtc1}). With \mlstat, IFAR value of the most significant foreground event GW151012 (GW170729) improves from $5.84$ $(0.73)$ years to $218.1$ $(55.8)$ years. We also observed that the event sequence according to IFAR values was considerably shuffled in \mlstat. The second most significant event viz., 151016, in O1 chunk has IFAR $0.21$ years.}
    \label{fig:sign_comp}
\end{figure*}

\begin{figure}[htb]
  \centering
  \includegraphics[width=\linewidth]{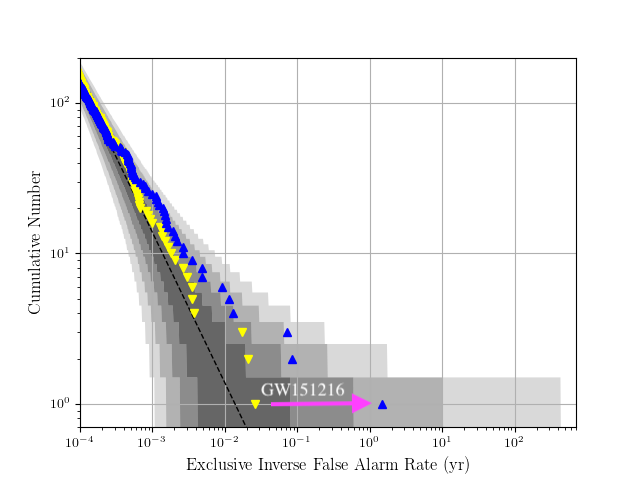}%
  \caption{Significance improvement for \newevent: With \mlstat, IFAR value of \newevent improves from $0.0173$ years to $1.453$ years making it the most significant foreground event of the chunk.}
    \label{fig:sign_comp_new_event}
\end{figure}

\begin{figure}[htb]
  \centering
  \includegraphics[width=\linewidth]{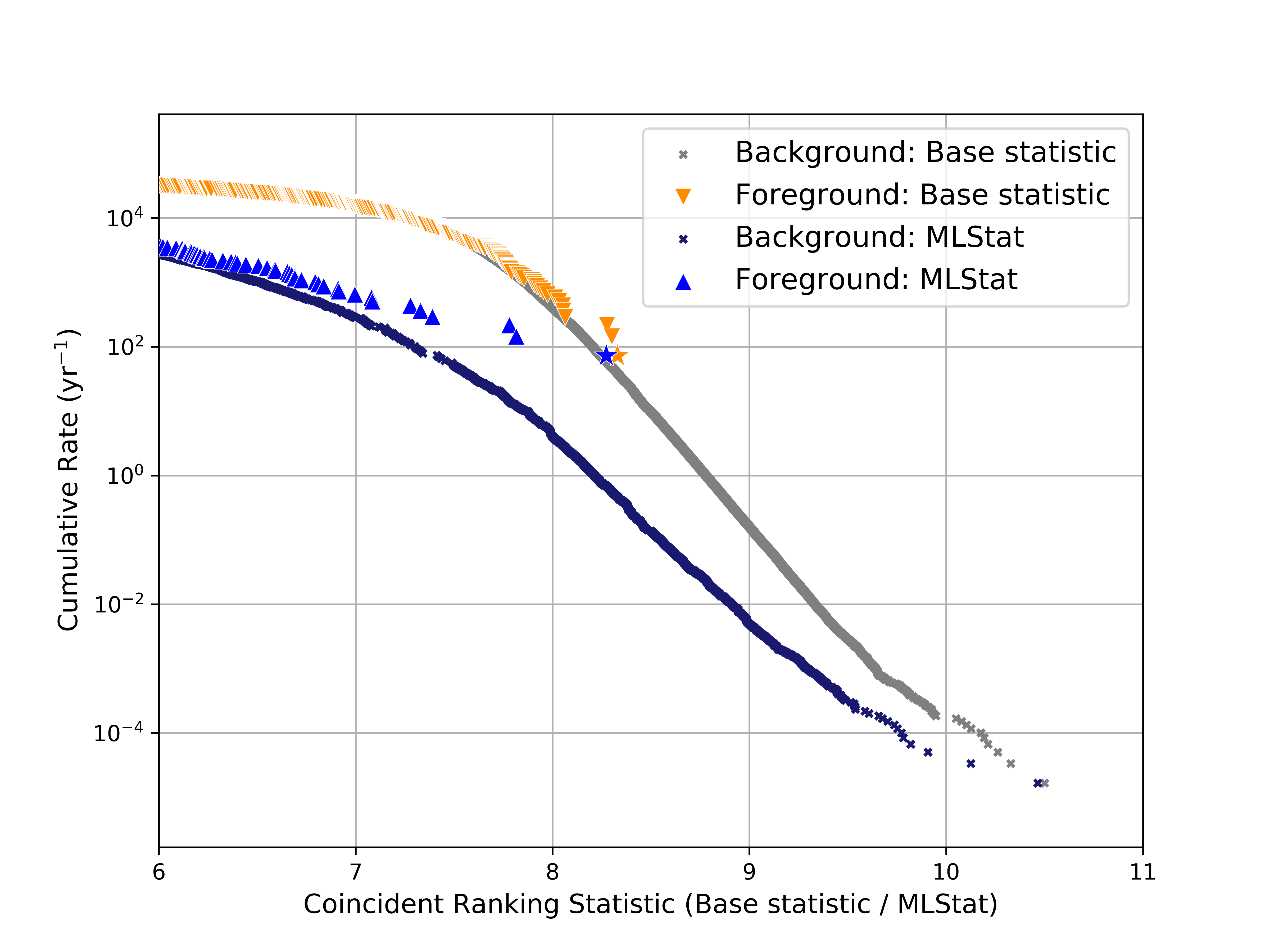}%
  \caption{Comparative search results with \mlstat and base statistic for the chunk containing \newevent (marked by star). Notice the reduced background and a better separated foreground with \mlstat.}
    \label{fig:frgbkg_new_event}
\end{figure}

We re-purpose the offline analysis data of \pycbc search described in GWTC-1~\cite{gwtc1, canton2017designing}. Initially, we analysed two chunks of data from O1 and O2 that contained the low significance events GW151012 and GW170729 respectively. The analysis consisted of $\sim5.9$ days of coincident data during October 8-20, 2015 for O1 and $\sim5.3$ days of coincident data starting from July 27 to Aug 5, 2017 for O2. As described in the next section, after finding the event \newevent with an improved estimated significance, the corresponding chunk consisting of $\sim5.1$ days of coincident data during December 3-18, 2015 was also analysed.

CWT maps of duration $1$ second were created keeping the \pycbc triggers in the centre. These images were then analysed with the ML classifier to get the respective \Pcbc values. We observed that the classifier is immune to changes in CWT maps corresponding to small translations in time, thus allowing us to round off the trigger GPS times to the first decimal place. Analysing multiple triggers lying within a time window of $0.1$ second is thus avoided. The \Pcbc values were recorded for all the triggers from both the detectors and the coincident analysis was carried out with \mlstat. The improvement in significance of GW151012 and GW170729 with \mlstat is shown in Fig.~\ref{fig:sign_comp}. The IFAR of GW151012 (GW170729) increases from $5.84$ $(0.73)$ years in base statistic to $218.1$ $(55.8)$ years in \mlstat, thus making them very confident detections. 

\subsection{Estimation for full foreground}\label{sec:ext_analysis}

\begin{figure}
  \centering
  \includegraphics[width=\linewidth]{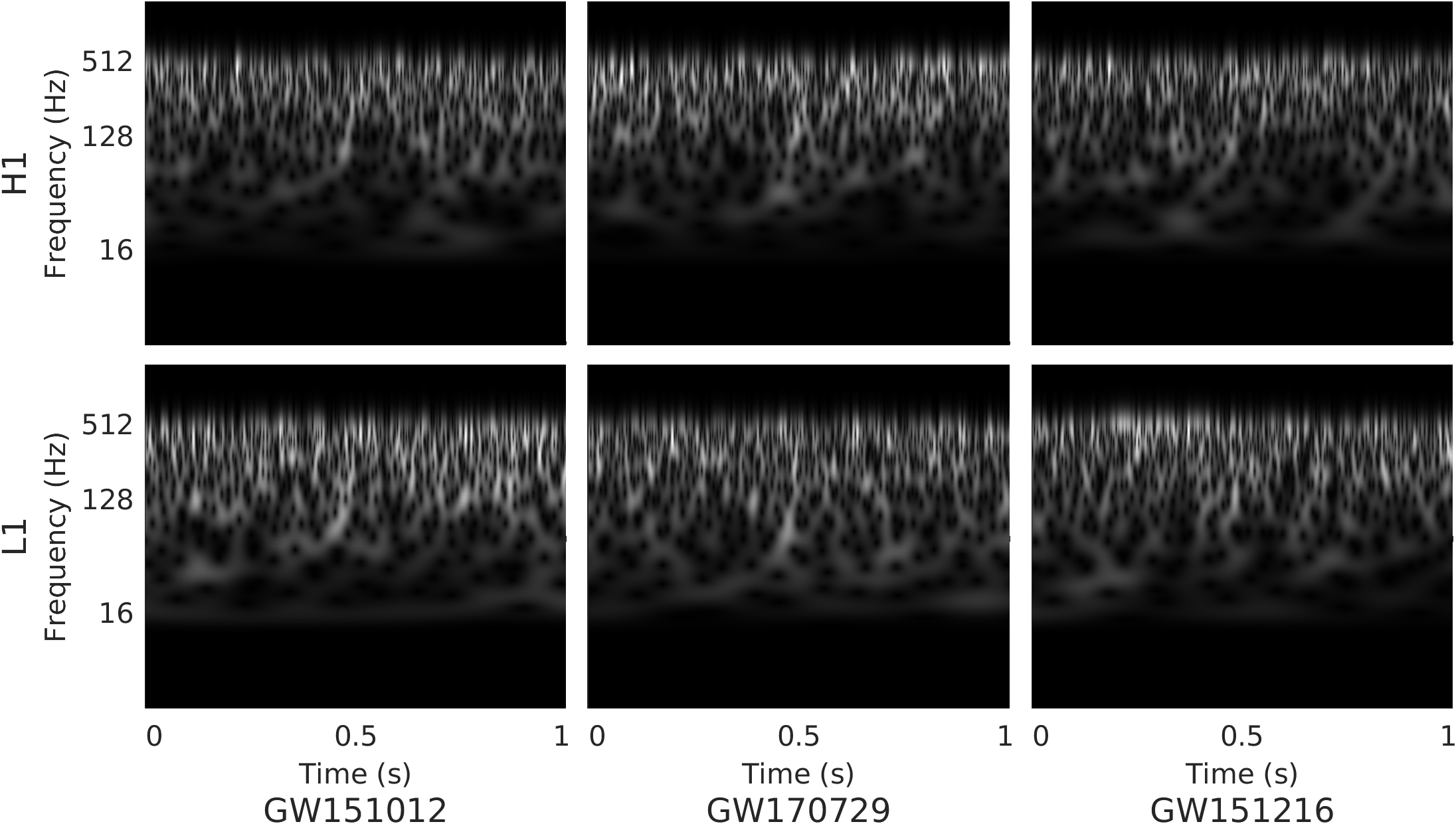}%
  \caption{CWT maps of events that gain significance with \mlstat. Whitened strain data of duration one second, band-passed between $16$~Hz and $512$~Hz, is used to generate the grayscale images that are fed to the classifier.}
  \label{fig:CWT_new_events}
\end{figure}

Though doing a full analysis of any of the observational runs is beyond the scope of the current work, we wish to get an estimate of what we should expect from the extended analysis. We make an assumption that the factor by which the background reduces as a function of base statistic remains similar across all the chunks in an observational run. We then note the improvement in the significance of the full set of foreground triggers against the background of the analysed chunk of that run and estimate their improved IFAR values with respect to the background of the original chunks they belonged to. Calculating these \mlstat IFAR values of full O1 and O2 foregrounds, we report the event \newevent in O1 with improved significance, which has been discussed in the literature before~\cite{2OGC, IAS_O1, IAS_GW151216, odds_GW151216, pratten2020assessing, huang2020source}. The IFAR for this event improved from $0.0173$ years in GWTC-1 to $1.656$ years. When the chunk containing the event was analysed with \mlstat, the actual IFAR value was obtained to be $1.453$ years, remarkably close to the one predicted using the approximate projection mentioned above (see Fig.~\ref{fig:sign_comp_new_event}). This confirmed the detection of \newevent with \mlstat while also increasing the reliability of the approximation. 
The comparison of foreground and background triggers with \mlstat and base statistic is shown in Fig.~\ref{fig:frgbkg_new_event}. Notice that the background and most of the foreground gets pushed to lower values with \mlstat except the event \newevent (marked with star) which shows minimal change in statistic value due to the high \Pcbc values recorded in both the detectors. The CWT maps of the events that become significant with our analysis are shown in Fig.~\ref{fig:CWT_new_events}.

\begin{figure}
\includegraphics[width=\columnwidth]{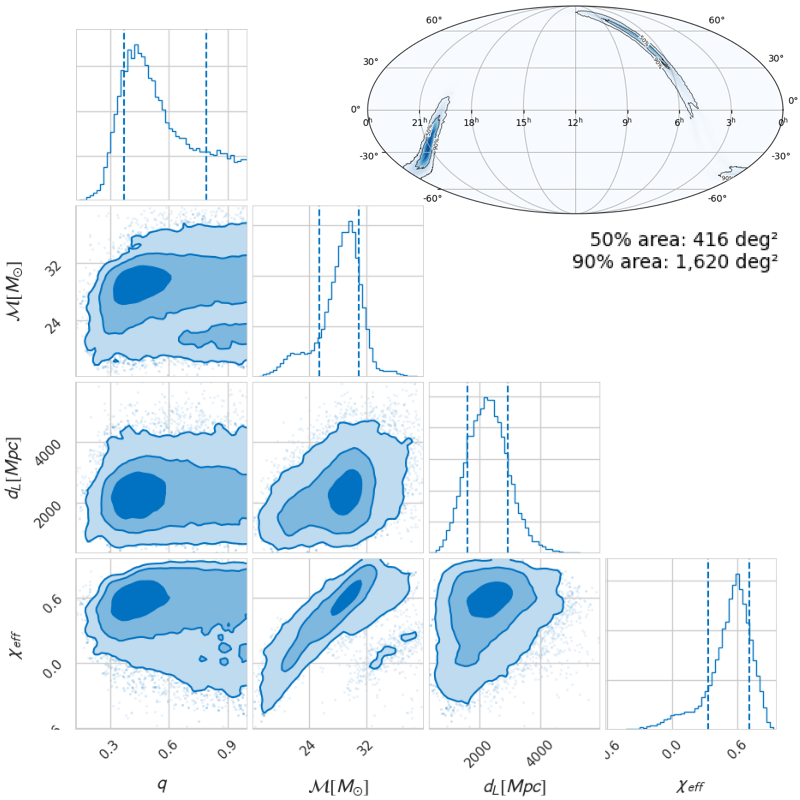}
\caption{Bayesian parameter estimation posteriors for the event \newevent using \waveform waveform model. We can clearly see the evidence for unequal masses with component masses of $\massone$ and $\masstwo$ and non-zero spin with $\chieff$ of $\chieffval$ and luminosity distance of $\lumdist$. The corresponding sky localisation map is also overlaid.}
\label{fig:PE}
\end{figure}

We list the combined foreground of O1 and O2 with $\mathrm{FAR}_\mathrm{ml}< 1$ per month in Table~\ref{tab:All_events}. GW170818 was not detected by \pycbc and thus does not show up in our analysis. The recorded \Pcbc values for GW170817 are very low in both the detectors as it falls out of the parameter space used for training the classifier and the CBC-tracks for such long duration signals may not be visible in the CWT map. However, as mentioned towards the end of Section~\ref{sec:construct_mlstat}, the loudness of this event results in \mlstat being only slightly smaller than the base statistic. We intend to build a more comprehensive ML tool better covering the parameter space of CBCs to include neutron stars as well in the followup work.

We report the parameter estimation results for \newevent using the fully Bayesian code \bilby~\cite{Ashton:2018jfp, bilby_gwtc1}. For the analysis, we estimate the noise power spectral density with \bayeswave~\cite{Becsy:2016ofp} using $16$~sec of data around \newevent. The posteriors with \waveform model~\cite{Cotesta:2020qhw} which includes higher order modes, are shown in the Fig.~\ref{fig:PE} for mass ratio, chirp mass, luminosity distance and effective spin. Other detailed analyses for the event which take into account various waveform models and priors can be found in~\cite{odds_GW151216, pratten2020assessing, huang2020source}. We find that \newevent has component masses of $\massone$ and $\masstwo$ with the inverse mass ratio of $\sim 2$ with the luminosity distance of $\lumdist$. Also 90\% credible intervals for effective inspiral spin parameter show non-zero spins with $\chieff$ of $\chieffval$. The expected posteriors from the parameter estimation also serves as a check that the event is indeed a BBH merger. We also plan to analyse the other marginal events in detail with data quality checks and parameter estimation.

\section{Conclusions}\label{sec:conclusion}

As the sensitivities of ground-based GW interferometers improve, the rate of detection of astrophysical events is going to increase, posing a challenge to the analyses to cope up with the large number of events. The major hurdle in this task arises from the  terrestrial and instrumental glitches, and their occurrence may increase with sensitivity. This issue may become even more severe with time as the density of events observed in the distant universe with statistically lower SNRs is expected to increase. Improving the fraction of true events in the set of potential triggers can significantly reduce this burden. Perhaps in the future ML will also help us in expanding the dimension and volume of the parameter space for astrophysical searches.

In this work, we demonstrated the capability of machine learning to improve the significance of CBC signals and to discard false triggers by integrating it with the existing analysis framework of \pycbc. We used transfer learning with InceptionV3, a pre-trained image based classifier, for effective identification of binary black hole mergers against glitches in LIGO data. We repurpose the \pycbc offline search data to re-analyse the matched filter triggers for two chunks of data from O1 and O2 that contained the events GW151012 and GW170729. We use the retrained InceptionV3 network to classify the continuous wavelet transform maps, a representation of time series data in time-frequency domain, of these triggers and get the \Pcbc values which are used to construct a new ranking statistic \mlstat -- a simple extension of the statistic used by \pycbc in the first GW transient catalog (GWTC-1). This helped in breaking the degeneracy between the real CBC signals and the noise transients that result in an increased background. For validating the performance of \mlstat against the base statistic, we performed a rigorous injection study with simulated BNS and BBH signals which showed on the average $\sim 10\%$ increase in sensitive volume for chirp masses between $0.8-5~\msun$ and $\sim 30\%$ for chirp masses higher than $5 \msun$. With the present training range being restricted to and data pre-processing favourable for the stellar mass BBHs (m$_1$,m$_2$ $\in 2-98\msun$; M $\leq 100\msun$), the ML network is not expected to identify signals in the low chirp masses. However, due to aggressive reduction in overall background with \mlstat, the detection sensitivity for low chirp mass signals is still enhanced. The sensitivity across the parameter space should further improve once we incorporate binaries involving neutron stars and intermediate mass BBHs in our training set. The pre-classification information may also be fed to the ML model to make its performance more robust.

With \mlstat, we achieve a considerable reduction in background and an improved separation of foreground. We report one to two orders of magnitude reduction in false alarm rates for the low significance events GW151012 and GW170729 with \mlstat as compared to the values obtained by \pycbc in GWTC-1. This is also the first time a machine learning based search algorithm was able to detect all the CBCs in GWTC-1 with same or better significance. We also confirm the detection of the event \newevent, which was not included in GWTC-1 lists of confirmed and marginal events due to lack of significance. While the existence and nature of this event is debated~\cite{IAS_GW151216, 2OGC, odds_GW151216, pratten2020assessing, huang2020source}, for demonstration of completeness of the process of obtaining a new detection using \mlstat, we perform parameter estimation for this event with an improved waveform model with higher order modes \waveform~\cite{Cotesta:2020qhw}. It is worth noting that the posteriors of parameters and the sky localisation contours do not suggest any obvious presence of a glitch and thus, strengthen the possibility of \newevent being of astrophysical origin.
Considering the ability of our method to distinguish the false triggers from astrophysical events, the list of sub-threshold events reported with \mlstat may be more reliable for astrophysics (e.g., population studies) but further follow-up through a data quality check would still be required.

That, by tuning a generic ML algorithm and introducing a simple extension to the ranking statistic, we could achieve these significant improvements, shows the enormous potential in ML, provided we can adapt it with fine understanding of the problem in hand. There is ample scope to improve our present analysis and, in general, several avenues may be explored to introduce machine learning based algorithms in GW analyses.

\begin{acknowledgments}
Authors express thanks to Ninan Sajeeth Philip, Shivaraj Kandhasamy and the LIGO-Virgo-KAGRA Collaboration for their valuable comments and suggestions. Some of the results in this work have been derived using the {\sc pesummary}
package~\cite{Hoy:2020vys}. SJ and NM acknowledge support of Council for Scientific and Industrial Research (CSIR), India. NM expresses thanks to the Max Planck Society and the Leibnitz University, Hannover. BG acknowledges support of the Max Planck Society and the University Grants Commission (UGC), India. We acknowledge the use of GWA and the LDG clusters at The Inter-University Centre for Astronomy and Astrophysics (Sarathi) and Caltech for the computational work. The follow up analysis for \newevent is performed with the HPC clusters {\sc Hypatia} at the Max Planck Institute  for  Gravitational  Physics, Potsdam-Golm. SM and SA acknowledge support from the Department of Science and Technology (DST), India, provided under the Swarna Jayanti Fellowships scheme. This document has been assigned \mbox{IUCAA} preprint number IUCAA-03/2020 and LIGO document number LIGO-P2000399.
\end{acknowledgments}

\bibliography{main}

\begin{thebibliography}{88}%
\makeatletter
\providecommand \@ifxundefined [1]{%
 \@ifx{#1\undefined}
}%
\providecommand \@ifnum [1]{%
 \ifnum #1\expandafter \@firstoftwo
 \else \expandafter \@secondoftwo
 \fi
}%
\providecommand \@ifx [1]{%
 \ifx #1\expandafter \@firstoftwo
 \else \expandafter \@secondoftwo
 \fi
}%
\providecommand \natexlab [1]{#1}%
\providecommand \enquote  [1]{``#1''}%
\providecommand \bibnamefont  [1]{#1}%
\providecommand \bibfnamefont [1]{#1}%
\providecommand \citenamefont [1]{#1}%
\providecommand \href@noop [0]{\@secondoftwo}%
\providecommand \href [0]{\begingroup \@sanitize@url \@href}%
\providecommand \@href[1]{\@@startlink{#1}\@@href}%
\providecommand \@@href[1]{\endgroup#1\@@endlink}%
\providecommand \@sanitize@url [0]{\catcode `\\12\catcode `\$12\catcode
  `\&12\catcode `\#12\catcode `\^12\catcode `\_12\catcode `\%12\relax}%
\providecommand \@@startlink[1]{}%
\providecommand \@@endlink[0]{}%
\providecommand \url  [0]{\begingroup\@sanitize@url \@url }%
\providecommand \@url [1]{\endgroup\@href {#1}{\urlprefix }}%
\providecommand \urlprefix  [0]{URL }%
\providecommand \Eprint [0]{\href }%
\providecommand \doibase [0]{http://dx.doi.org/}%
\providecommand \selectlanguage [0]{\@gobble}%
\providecommand \bibinfo  [0]{\@secondoftwo}%
\providecommand \bibfield  [0]{\@secondoftwo}%
\providecommand \translation [1]{[#1]}%
\providecommand \BibitemOpen [0]{}%
\providecommand \bibitemStop [0]{}%
\providecommand \bibitemNoStop [0]{.\EOS\space}%
\providecommand \EOS [0]{\spacefactor3000\relax}%
\providecommand \BibitemShut  [1]{\csname bibitem#1\endcsname}%
\let\auto@bib@innerbib\@empty
\bibitem [{\citenamefont {Aasi}\ \emph {et~al.}(2015)\citenamefont {Aasi} \emph
  {et~al.}}]{aLIGO}%
  \BibitemOpen
  \bibfield  {author} {\bibinfo {author} {\bibfnamefont {J.}~\bibnamefont
  {Aasi}} \emph {et~al.} (\bibinfo {collaboration} {LIGO Scientific
  Collaboration and Virgo Collaboration}),\ }\href {\doibase
  10.1088/0264-9381/32/7/074001} {\bibfield  {journal} {\bibinfo  {journal}
  {Classical and Quantum Gravity}\ }\textbf {\bibinfo {volume} {32}},\ \bibinfo
  {pages} {074001} (\bibinfo {year} {2015})}\BibitemShut {NoStop}%
\bibitem [{\citenamefont {Abbott}\ \emph
  {et~al.}(2016{\natexlab{a}})\citenamefont {Abbott} \emph
  {et~al.}}]{aLIGO_Detectors}%
  \BibitemOpen
  \bibfield  {author} {\bibinfo {author} {\bibfnamefont {B.~P.}\ \bibnamefont
  {Abbott}} \emph {et~al.} (\bibinfo {collaboration} {LIGO Scientific
  Collaboration and Virgo Collaboration}),\ }\href {\doibase
  10.1103/PhysRevLett.116.131103} {\bibfield  {journal} {\bibinfo  {journal}
  {Phys. Rev. Lett.}\ }\textbf {\bibinfo {volume} {116}},\ \bibinfo {pages}
  {131103} (\bibinfo {year} {2016}{\natexlab{a}})}\BibitemShut {NoStop}%
\bibitem [{\citenamefont {Acernese}\ \emph {et~al.}(2015)\citenamefont
  {Acernese} \emph {et~al.}}]{aVIRGO}%
  \BibitemOpen
  \bibfield  {author} {\bibinfo {author} {\bibfnamefont {F.}~\bibnamefont
  {Acernese}} \emph {et~al.},\ }\href {\doibase 10.1088/0264-9381/32/2/024001}
  {\bibfield  {journal} {\bibinfo  {journal} {Classical and Quantum Gravity}\
  }\textbf {\bibinfo {volume} {32}},\ \bibinfo {pages} {024001} (\bibinfo
  {year} {2015})}\BibitemShut {NoStop}%
\bibitem [{\citenamefont {Abbott}\ \emph
  {et~al.}(2016{\natexlab{b}})\citenamefont {Abbott} \emph
  {et~al.}}]{GW150914}%
  \BibitemOpen
  \bibfield  {author} {\bibinfo {author} {\bibfnamefont {B.~P.}\ \bibnamefont
  {Abbott}} \emph {et~al.} (\bibinfo {collaboration} {LIGO Scientific
  Collaboration and Virgo Collaboration}),\ }\href {\doibase
  10.1103/PhysRevLett.116.061102} {\bibfield  {journal} {\bibinfo  {journal}
  {Phys. Rev. Lett.}\ }\textbf {\bibinfo {volume} {116}},\ \bibinfo {pages}
  {061102} (\bibinfo {year} {2016}{\natexlab{b}})}\BibitemShut {NoStop}%
\bibitem [{\citenamefont {Abbott}\ \emph
  {et~al.}(2016{\natexlab{c}})\citenamefont {Abbott} \emph
  {et~al.}}]{GW151226}%
  \BibitemOpen
  \bibfield  {author} {\bibinfo {author} {\bibfnamefont {B.~P.}\ \bibnamefont
  {Abbott}} \emph {et~al.} (\bibinfo {collaboration} {LIGO Scientific
  Collaboration and Virgo Collaboration}),\ }\href {\doibase
  10.1103/PhysRevLett.116.241103} {\bibfield  {journal} {\bibinfo  {journal}
  {Phys. Rev. Lett.}\ }\textbf {\bibinfo {volume} {116}},\ \bibinfo {pages}
  {241103} (\bibinfo {year} {2016}{\natexlab{c}})}\BibitemShut {NoStop}%
\bibitem [{\citenamefont {Abbott}\ \emph
  {et~al.}(2017{\natexlab{a}})\citenamefont {Abbott} \emph
  {et~al.}}]{GW170104}%
  \BibitemOpen
  \bibfield  {author} {\bibinfo {author} {\bibfnamefont {B.~P.}\ \bibnamefont
  {Abbott}} \emph {et~al.} (\bibinfo {collaboration} {LIGO Scientific and Virgo
  Collaboration}),\ }\href {\doibase 10.1103/PhysRevLett.118.221101} {\bibfield
   {journal} {\bibinfo  {journal} {Phys. Rev. Lett.}\ }\textbf {\bibinfo
  {volume} {118}},\ \bibinfo {pages} {221101} (\bibinfo {year}
  {2017}{\natexlab{a}})}\BibitemShut {NoStop}%
\bibitem [{\citenamefont {Abbott}\ \emph
  {et~al.}(2017{\natexlab{b}})\citenamefont {Abbott} \emph
  {et~al.}}]{GW170608}%
  \BibitemOpen
  \bibfield  {author} {\bibinfo {author} {\bibfnamefont {B.~P.}\ \bibnamefont
  {Abbott}} \emph {et~al.},\ }\href {\doibase 10.3847/2041-8213/aa9f0c}
  {\bibfield  {journal} {\bibinfo  {journal} {The Astrophysical Journal}\
  }\textbf {\bibinfo {volume} {851}},\ \bibinfo {pages} {L35} (\bibinfo {year}
  {2017}{\natexlab{b}})}\BibitemShut {NoStop}%
\bibitem [{\citenamefont {Abbott}\ \emph
  {et~al.}(2017{\natexlab{c}})\citenamefont {Abbott} \emph
  {et~al.}}]{GW170814}%
  \BibitemOpen
  \bibfield  {author} {\bibinfo {author} {\bibfnamefont {B.~P.}\ \bibnamefont
  {Abbott}} \emph {et~al.} (\bibinfo {collaboration} {LIGO Scientific
  Collaboration and Virgo Collaboration}),\ }\href {\doibase
  10.1103/PhysRevLett.119.141101} {\bibfield  {journal} {\bibinfo  {journal}
  {Phys. Rev. Lett.}\ }\textbf {\bibinfo {volume} {119}},\ \bibinfo {pages}
  {141101} (\bibinfo {year} {2017}{\natexlab{c}})}\BibitemShut {NoStop}%
\bibitem [{\citenamefont {Abbott}\ \emph {et~al.}(2019)\citenamefont {Abbott}
  \emph {et~al.}}]{gwtc1}%
  \BibitemOpen
  \bibfield  {author} {\bibinfo {author} {\bibfnamefont {B.~P.}\ \bibnamefont
  {Abbott}} \emph {et~al.} (\bibinfo {collaboration} {LIGO Scientific
  Collaboration and Virgo Collaboration}),\ }\href {\doibase
  10.1103/PhysRevX.9.031040} {\bibfield  {journal} {\bibinfo  {journal} {Phys.
  Rev. X}\ }\textbf {\bibinfo {volume} {9}},\ \bibinfo {pages} {031040}
  (\bibinfo {year} {2019})}\BibitemShut {NoStop}%
\bibitem [{\citenamefont {Abbott}\ \emph {et~al.}(2021)\citenamefont {Abbott}
  \emph {et~al.}}]{gwtc2}%
  \BibitemOpen
  \bibfield  {author} {\bibinfo {author} {\bibfnamefont {R.}~\bibnamefont
  {Abbott}} \emph {et~al.},\ }\href {\doibase 10.1103/physrevx.11.021053}
  {\bibfield  {journal} {\bibinfo  {journal} {Physical Review X}\ }\textbf
  {\bibinfo {volume} {11}} (\bibinfo {year} {2021}),\
  10.1103/physrevx.11.021053}\BibitemShut {NoStop}%
\bibitem [{\citenamefont {Abbott}\ \emph
  {et~al.}(2020{\natexlab{a}})\citenamefont {Abbott} \emph
  {et~al.}}]{GW190412}%
  \BibitemOpen
  \bibfield  {author} {\bibinfo {author} {\bibfnamefont {R.}~\bibnamefont
  {Abbott}} \emph {et~al.} (\bibinfo {collaboration} {LIGO Scientific
  Collaboration and Virgo Collaboration}),\ }\href {\doibase
  10.1103/PhysRevD.102.043015} {\bibfield  {journal} {\bibinfo  {journal}
  {Phys. Rev. D}\ }\textbf {\bibinfo {volume} {102}},\ \bibinfo {pages}
  {043015} (\bibinfo {year} {2020}{\natexlab{a}})}\BibitemShut {NoStop}%
\bibitem [{\citenamefont {Abbott}\ \emph
  {et~al.}(2020{\natexlab{b}})\citenamefont {Abbott} \emph
  {et~al.}}]{GW190521}%
  \BibitemOpen
  \bibfield  {author} {\bibinfo {author} {\bibfnamefont {R.}~\bibnamefont
  {Abbott}} \emph {et~al.} (\bibinfo {collaboration} {LIGO Scientific
  Collaboration and Virgo Collaboration}),\ }\href {\doibase
  10.1103/PhysRevLett.125.101102} {\bibfield  {journal} {\bibinfo  {journal}
  {Phys. Rev. Lett.}\ }\textbf {\bibinfo {volume} {125}},\ \bibinfo {pages}
  {101102} (\bibinfo {year} {2020}{\natexlab{b}})}\BibitemShut {NoStop}%
\bibitem [{\citenamefont {Abbott}\ \emph
  {et~al.}(2020{\natexlab{c}})\citenamefont {Abbott} \emph
  {et~al.}}]{GW190814}%
  \BibitemOpen
  \bibfield  {author} {\bibinfo {author} {\bibfnamefont {R.}~\bibnamefont
  {Abbott}} \emph {et~al.},\ }\href {\doibase 10.3847/2041-8213/ab960f}
  {\bibfield  {journal} {\bibinfo  {journal} {The Astrophysical Journal}\
  }\textbf {\bibinfo {volume} {896}},\ \bibinfo {pages} {L44} (\bibinfo {year}
  {2020}{\natexlab{c}})}\BibitemShut {NoStop}%
\bibitem [{\citenamefont {Abbott}\ \emph
  {et~al.}(2017{\natexlab{d}})\citenamefont {Abbott} \emph
  {et~al.}}]{GW170817}%
  \BibitemOpen
  \bibfield  {author} {\bibinfo {author} {\bibfnamefont {B.~P.}\ \bibnamefont
  {Abbott}} \emph {et~al.} (\bibinfo {collaboration} {LIGO Scientific
  Collaboration and Virgo Collaboration}),\ }\href {\doibase
  10.1103/PhysRevLett.119.161101} {\bibfield  {journal} {\bibinfo  {journal}
  {Phys. Rev. Lett.}\ }\textbf {\bibinfo {volume} {119}},\ \bibinfo {pages}
  {161101} (\bibinfo {year} {2017}{\natexlab{d}})}\BibitemShut {NoStop}%
\bibitem [{\citenamefont {Abbott}\ \emph
  {et~al.}(2020{\natexlab{d}})\citenamefont {Abbott} \emph
  {et~al.}}]{GW190425}%
  \BibitemOpen
  \bibfield  {author} {\bibinfo {author} {\bibfnamefont {B.~P.}\ \bibnamefont
  {Abbott}} \emph {et~al.},\ }\href {\doibase 10.3847/2041-8213/ab75f5}
  {\bibfield  {journal} {\bibinfo  {journal} {The Astrophysical Journal}\
  }\textbf {\bibinfo {volume} {892}},\ \bibinfo {pages} {L3} (\bibinfo {year}
  {2020}{\natexlab{d}})}\BibitemShut {NoStop}%
\bibitem [{\citenamefont {Aso}\ \emph {et~al.}(2013)\citenamefont {Aso},
  \citenamefont {Michimura}, \citenamefont {Somiya}, \citenamefont {Ando},
  \citenamefont {Miyakawa}, \citenamefont {Sekiguchi}, \citenamefont
  {Tatsumi},\ and\ \citenamefont {Yamamoto}}]{Kagradesign}%
  \BibitemOpen
  \bibfield  {author} {\bibinfo {author} {\bibfnamefont {Y.}~\bibnamefont
  {Aso}}, \bibinfo {author} {\bibfnamefont {Y.}~\bibnamefont {Michimura}},
  \bibinfo {author} {\bibfnamefont {K.}~\bibnamefont {Somiya}}, \bibinfo
  {author} {\bibfnamefont {M.}~\bibnamefont {Ando}}, \bibinfo {author}
  {\bibfnamefont {O.}~\bibnamefont {Miyakawa}}, \bibinfo {author}
  {\bibfnamefont {T.}~\bibnamefont {Sekiguchi}}, \bibinfo {author}
  {\bibfnamefont {D.}~\bibnamefont {Tatsumi}}, \ and\ \bibinfo {author}
  {\bibfnamefont {H.}~\bibnamefont {Yamamoto}} (\bibinfo {collaboration} {The
  KAGRA Collaboration}),\ }\href {\doibase 10.1103/PhysRevD.88.043007}
  {\bibfield  {journal} {\bibinfo  {journal} {Phys. Rev. D}\ }\textbf {\bibinfo
  {volume} {88}},\ \bibinfo {pages} {043007} (\bibinfo {year}
  {2013})}\BibitemShut {NoStop}%
\bibitem [{\citenamefont {Akutsu}\ \emph {et~al.}(2020)\citenamefont {Akutsu},
  \citenamefont {Ando}, \citenamefont {Arai}, \citenamefont {Arai},
  \citenamefont {Araki}, \citenamefont {Araya} \emph
  {et~al.}}]{akutsu2020overview}%
  \BibitemOpen
  \bibfield  {author} {\bibinfo {author} {\bibfnamefont {T.}~\bibnamefont
  {Akutsu}}, \bibinfo {author} {\bibfnamefont {M.}~\bibnamefont {Ando}},
  \bibinfo {author} {\bibfnamefont {K.}~\bibnamefont {Arai}}, \bibinfo {author}
  {\bibfnamefont {Y.}~\bibnamefont {Arai}}, \bibinfo {author} {\bibfnamefont
  {S.}~\bibnamefont {Araki}}, \bibinfo {author} {\bibfnamefont
  {A.}~\bibnamefont {Araya}},  \emph {et~al.},\ }\href@noop {} {\enquote
  {\bibinfo {title} {Overview of kagra: Detector design and construction
  history},}\ } (\bibinfo {year} {2020}),\ \Eprint
  {http://arxiv.org/abs/2005.05574} {arXiv:2005.05574 [physics.ins-det]}
  \BibitemShut {NoStop}%
\bibitem [{\citenamefont {Abbott}\ \emph {et~al.}(2018)\citenamefont {Abbott}
  \emph {et~al.}}]{abbott2018prospects}%
  \BibitemOpen
  \bibfield  {author} {\bibinfo {author} {\bibfnamefont {B.~P.}\ \bibnamefont
  {Abbott}} \emph {et~al.},\ }\href
  {https://doi.org/10.1007/s41114-020-00026-9} {\bibfield  {journal} {\bibinfo
  {journal} {Living Reviews in Relativity}\ }\textbf {\bibinfo {volume} {21}},\
  \bibinfo {pages} {3} (\bibinfo {year} {2018})}\BibitemShut {NoStop}%
\bibitem [{\citenamefont {Usman}\ \emph {et~al.}(2016)\citenamefont {Usman},
  \citenamefont {Nitz}, \citenamefont {Harry}, \citenamefont {Biwer},
  \citenamefont {Brown} \emph {et~al.}}]{Usman_2016}%
  \BibitemOpen
  \bibfield  {author} {\bibinfo {author} {\bibfnamefont {S.~A.}\ \bibnamefont
  {Usman}}, \bibinfo {author} {\bibfnamefont {A.~H.}\ \bibnamefont {Nitz}},
  \bibinfo {author} {\bibfnamefont {I.~W.}\ \bibnamefont {Harry}}, \bibinfo
  {author} {\bibfnamefont {C.~M.}\ \bibnamefont {Biwer}}, \bibinfo {author}
  {\bibfnamefont {D.~A.}\ \bibnamefont {Brown}},  \emph {et~al.},\ }\href
  {\doibase 10.1088/0264-9381/33/21/215004} {\bibfield  {journal} {\bibinfo
  {journal} {Classical and Quantum Gravity}\ }\textbf {\bibinfo {volume}
  {33}},\ \bibinfo {pages} {215004} (\bibinfo {year} {2016})}\BibitemShut
  {NoStop}%
\bibitem [{\citenamefont {Sachdev}\ \emph {et~al.}(2019)\citenamefont
  {Sachdev}, \citenamefont {Caudill}, \citenamefont {Fong}, \citenamefont {Lo},
  \citenamefont {Messick} \emph {et~al.}}]{sachdev2019gstlal}%
  \BibitemOpen
  \bibfield  {author} {\bibinfo {author} {\bibfnamefont {S.}~\bibnamefont
  {Sachdev}}, \bibinfo {author} {\bibfnamefont {S.}~\bibnamefont {Caudill}},
  \bibinfo {author} {\bibfnamefont {H.}~\bibnamefont {Fong}}, \bibinfo {author}
  {\bibfnamefont {R.~K.~L.}\ \bibnamefont {Lo}}, \bibinfo {author}
  {\bibfnamefont {C.}~\bibnamefont {Messick}},  \emph {et~al.},\ }\href@noop {}
  {\enquote {\bibinfo {title} {The gstlal search analysis methods for compact
  binary mergers in advanced ligo's second and advanced virgo's first observing
  runs},}\ } (\bibinfo {year} {2019}),\ \Eprint
  {http://arxiv.org/abs/1901.08580} {arXiv:1901.08580 [gr-qc]} \BibitemShut
  {NoStop}%
\bibitem [{\citenamefont {Klimenko}\ \emph {et~al.}(2016)\citenamefont
  {Klimenko}, \citenamefont {Vedovato}, \citenamefont {Drago}, \citenamefont
  {Salemi}, \citenamefont {Tiwari} \emph {et~al.}}]{klimenko2016method}%
  \BibitemOpen
  \bibfield  {author} {\bibinfo {author} {\bibfnamefont {S.}~\bibnamefont
  {Klimenko}}, \bibinfo {author} {\bibfnamefont {G.}~\bibnamefont {Vedovato}},
  \bibinfo {author} {\bibfnamefont {M.}~\bibnamefont {Drago}}, \bibinfo
  {author} {\bibfnamefont {F.}~\bibnamefont {Salemi}}, \bibinfo {author}
  {\bibfnamefont {V.}~\bibnamefont {Tiwari}},  \emph {et~al.},\ }\href
  {\doibase 10.1103/PhysRevD.93.042004} {\bibfield  {journal} {\bibinfo
  {journal} {Phys. Rev. D}\ }\textbf {\bibinfo {volume} {93}},\ \bibinfo
  {pages} {042004} (\bibinfo {year} {2016})}\BibitemShut {NoStop}%
\bibitem [{\citenamefont {Adams}(2015)}]{adams2015low}%
  \BibitemOpen
  \bibfield  {author} {\bibinfo {author} {\bibfnamefont {T.}~\bibnamefont
  {Adams}} (\bibinfo {collaboration} {LIGO Scientific Collaboration and Virgo
  Collaboration})\ }(\bibinfo {year} {2015})\ \Eprint
  {http://arxiv.org/abs/1507.01787} {arXiv:1507.01787 [gr-qc]} \BibitemShut
  {NoStop}%
\bibitem [{\citenamefont {{Chu}}\ \emph {et~al.}(2020)\citenamefont {{Chu}},
  \citenamefont {{Kovalam}}, \citenamefont {{Wen}}, \citenamefont
  {{Slaven-Blair}}, \citenamefont {{Bosveld}}, \citenamefont {{Chen}},
  \citenamefont {{Clearwater}}, \citenamefont {{Codoreanu}}, \citenamefont
  {{Du}}, \citenamefont {{Guo}}, \citenamefont {{Guo}}, \citenamefont {{Kim}},
  \citenamefont {{Li}}, \citenamefont {{Oloworaran}}, \citenamefont
  {{Panther}}, \citenamefont {{Powell}}, \citenamefont {{Sengupta}},
  \citenamefont {{Wette}},\ and\ \citenamefont {{Zhu}}}]{2020arXiv201106787C}%
  \BibitemOpen
  \bibfield  {author} {\bibinfo {author} {\bibfnamefont {Q.}~\bibnamefont
  {{Chu}}}, \bibinfo {author} {\bibfnamefont {M.}~\bibnamefont {{Kovalam}}},
  \bibinfo {author} {\bibfnamefont {L.}~\bibnamefont {{Wen}}}, \bibinfo
  {author} {\bibfnamefont {T.}~\bibnamefont {{Slaven-Blair}}}, \bibinfo
  {author} {\bibfnamefont {J.}~\bibnamefont {{Bosveld}}}, \bibinfo {author}
  {\bibfnamefont {Y.}~\bibnamefont {{Chen}}}, \bibinfo {author} {\bibfnamefont
  {P.}~\bibnamefont {{Clearwater}}}, \bibinfo {author} {\bibfnamefont
  {A.}~\bibnamefont {{Codoreanu}}}, \bibinfo {author} {\bibfnamefont
  {Z.}~\bibnamefont {{Du}}}, \bibinfo {author} {\bibfnamefont {X.}~\bibnamefont
  {{Guo}}}, \bibinfo {author} {\bibfnamefont {X.}~\bibnamefont {{Guo}}},
  \bibinfo {author} {\bibfnamefont {K.}~\bibnamefont {{Kim}}}, \bibinfo
  {author} {\bibfnamefont {T.~G.~F.}\ \bibnamefont {{Li}}}, \bibinfo {author}
  {\bibfnamefont {V.}~\bibnamefont {{Oloworaran}}}, \bibinfo {author}
  {\bibfnamefont {F.}~\bibnamefont {{Panther}}}, \bibinfo {author}
  {\bibfnamefont {J.}~\bibnamefont {{Powell}}}, \bibinfo {author}
  {\bibfnamefont {A.~S.}\ \bibnamefont {{Sengupta}}}, \bibinfo {author}
  {\bibfnamefont {K.}~\bibnamefont {{Wette}}}, \ and\ \bibinfo {author}
  {\bibfnamefont {X.}~\bibnamefont {{Zhu}}},\ }\href@noop {} {\bibfield
  {journal} {\bibinfo  {journal} {arXiv e-prints}\ ,\ \bibinfo {eid}
  {arXiv:2011.06787}} (\bibinfo {year} {2020})},\ \Eprint
  {http://arxiv.org/abs/2011.06787} {arXiv:2011.06787 [gr-qc]} \BibitemShut
  {NoStop}%
\bibitem [{\citenamefont {Zubakov}\ and\ \citenamefont
  {Wainstein}(1962)}]{wainsteinextraction}%
  \BibitemOpen
  \bibfield  {author} {\bibinfo {author} {\bibfnamefont {L.}~\bibnamefont
  {Zubakov}}\ and\ \bibinfo {author} {\bibfnamefont {V.}~\bibnamefont
  {Wainstein}},\ }\href@noop {} {\enquote {\bibinfo {title} {Extraction of
  signals from noise},}\ } (\bibinfo {year} {1962})\BibitemShut {NoStop}%
\bibitem [{\citenamefont {Cutler}\ \emph {et~al.}(1993)\citenamefont {Cutler},
  \citenamefont {Apostolatos}, \citenamefont {Bildsten}, \citenamefont {Finn},
  \citenamefont {Flanagan} \emph {et~al.}}]{Cutler}%
  \BibitemOpen
  \bibfield  {author} {\bibinfo {author} {\bibfnamefont {C.}~\bibnamefont
  {Cutler}}, \bibinfo {author} {\bibfnamefont {T.~A.}\ \bibnamefont
  {Apostolatos}}, \bibinfo {author} {\bibfnamefont {L.}~\bibnamefont
  {Bildsten}}, \bibinfo {author} {\bibfnamefont {L.~S.}\ \bibnamefont {Finn}},
  \bibinfo {author} {\bibfnamefont {E.~E.}\ \bibnamefont {Flanagan}},  \emph
  {et~al.},\ }\href {\doibase 10.1103/PhysRevLett.70.2984} {\bibfield
  {journal} {\bibinfo  {journal} {Phys. Rev. Lett.}\ }\textbf {\bibinfo
  {volume} {70}},\ \bibinfo {pages} {2984} (\bibinfo {year}
  {1993})}\BibitemShut {NoStop}%
\bibitem [{\citenamefont {Pai}\ \emph {et~al.}(2001)\citenamefont {Pai},
  \citenamefont {Dhurandhar},\ and\ \citenamefont {Bose}}]{Pai}%
  \BibitemOpen
  \bibfield  {author} {\bibinfo {author} {\bibfnamefont {A.}~\bibnamefont
  {Pai}}, \bibinfo {author} {\bibfnamefont {S.}~\bibnamefont {Dhurandhar}}, \
  and\ \bibinfo {author} {\bibfnamefont {S.}~\bibnamefont {Bose}},\ }\href
  {\doibase 10.1103/PhysRevD.64.042004} {\bibfield  {journal} {\bibinfo
  {journal} {Phys. Rev. D}\ }\textbf {\bibinfo {volume} {64}},\ \bibinfo
  {pages} {042004} (\bibinfo {year} {2001})}\BibitemShut {NoStop}%
\bibitem [{\citenamefont {Allen}\ \emph {et~al.}(2012)\citenamefont {Allen},
  \citenamefont {Anderson}, \citenamefont {Brady}, \citenamefont {Brown},\ and\
  \citenamefont {Creighton}}]{Allen12}%
  \BibitemOpen
  \bibfield  {author} {\bibinfo {author} {\bibfnamefont {B.}~\bibnamefont
  {Allen}}, \bibinfo {author} {\bibfnamefont {W.~G.}\ \bibnamefont {Anderson}},
  \bibinfo {author} {\bibfnamefont {P.~R.}\ \bibnamefont {Brady}}, \bibinfo
  {author} {\bibfnamefont {D.~A.}\ \bibnamefont {Brown}}, \ and\ \bibinfo
  {author} {\bibfnamefont {J.~D.~E.}\ \bibnamefont {Creighton}},\ }\href
  {\doibase 10.1103/PhysRevD.85.122006} {\bibfield  {journal} {\bibinfo
  {journal} {Phys. Rev. D}\ }\textbf {\bibinfo {volume} {85}},\ \bibinfo
  {pages} {122006} (\bibinfo {year} {2012})}\BibitemShut {NoStop}%
\bibitem [{\citenamefont {Finn}(1992)}]{Finn}%
  \BibitemOpen
  \bibfield  {author} {\bibinfo {author} {\bibfnamefont {L.~S.}\ \bibnamefont
  {Finn}},\ }\href {\doibase 10.1103/PhysRevD.46.5236} {\bibfield  {journal}
  {\bibinfo  {journal} {Phys. Rev. D}\ }\textbf {\bibinfo {volume} {46}},\
  \bibinfo {pages} {5236} (\bibinfo {year} {1992})}\BibitemShut {NoStop}%
\bibitem [{\citenamefont {Finn}\ and\ \citenamefont
  {Chernoff}(1993)}]{Finn_Chern}%
  \BibitemOpen
  \bibfield  {author} {\bibinfo {author} {\bibfnamefont {L.~S.}\ \bibnamefont
  {Finn}}\ and\ \bibinfo {author} {\bibfnamefont {D.~F.}\ \bibnamefont
  {Chernoff}},\ }\href {\doibase 10.1103/PhysRevD.47.2198} {\bibfield
  {journal} {\bibinfo  {journal} {Phys. Rev. D}\ }\textbf {\bibinfo {volume}
  {47}},\ \bibinfo {pages} {2198} (\bibinfo {year} {1993})}\BibitemShut
  {NoStop}%
\bibitem [{\citenamefont {Abbott}\ \emph
  {et~al.}(2016{\natexlab{d}})\citenamefont {Abbott} \emph
  {et~al.}}]{abbott2016characterization}%
  \BibitemOpen
  \bibfield  {author} {\bibinfo {author} {\bibfnamefont {B.~P.}\ \bibnamefont
  {Abbott}} \emph {et~al.},\ }\href {\doibase 10.1088/0264-9381/33/13/134001}
  {\bibfield  {journal} {\bibinfo  {journal} {Classical and Quantum Gravity}\
  }\textbf {\bibinfo {volume} {33}},\ \bibinfo {pages} {134001} (\bibinfo
  {year} {2016}{\natexlab{d}})}\BibitemShut {NoStop}%
\bibitem [{\citenamefont {Nuttall}\ \emph {et~al.}(2015)\citenamefont
  {Nuttall}, \citenamefont {Massinger}, \citenamefont {Areeda}, \citenamefont
  {Betzwieser}, \citenamefont {Dwyer}, \citenamefont {Effler}, \citenamefont
  {Fisher}, \citenamefont {Fritschel}, \citenamefont {Kissel}, \citenamefont
  {Lundgren}, \citenamefont {Macleod}, \citenamefont {Martynov}, \citenamefont
  {McIver}, \citenamefont {Mullavey}, \citenamefont {Sigg}, \citenamefont
  {Smith}, \citenamefont {Vajente}, \citenamefont {Williamson},\ and\
  \citenamefont {Wipf}}]{nuttall2015improving}%
  \BibitemOpen
  \bibfield  {author} {\bibinfo {author} {\bibfnamefont {L.~K.}\ \bibnamefont
  {Nuttall}}, \bibinfo {author} {\bibfnamefont {T.~J.}\ \bibnamefont
  {Massinger}}, \bibinfo {author} {\bibfnamefont {J.}~\bibnamefont {Areeda}},
  \bibinfo {author} {\bibfnamefont {J.}~\bibnamefont {Betzwieser}}, \bibinfo
  {author} {\bibfnamefont {S.}~\bibnamefont {Dwyer}}, \bibinfo {author}
  {\bibfnamefont {A.}~\bibnamefont {Effler}}, \bibinfo {author} {\bibfnamefont
  {R.~P.}\ \bibnamefont {Fisher}}, \bibinfo {author} {\bibfnamefont
  {P.}~\bibnamefont {Fritschel}}, \bibinfo {author} {\bibfnamefont {J.~S.}\
  \bibnamefont {Kissel}}, \bibinfo {author} {\bibfnamefont {A.~P.}\
  \bibnamefont {Lundgren}}, \bibinfo {author} {\bibfnamefont {D.~M.}\
  \bibnamefont {Macleod}}, \bibinfo {author} {\bibfnamefont {D.}~\bibnamefont
  {Martynov}}, \bibinfo {author} {\bibfnamefont {J.}~\bibnamefont {McIver}},
  \bibinfo {author} {\bibfnamefont {A.}~\bibnamefont {Mullavey}}, \bibinfo
  {author} {\bibfnamefont {D.}~\bibnamefont {Sigg}}, \bibinfo {author}
  {\bibfnamefont {J.~R.}\ \bibnamefont {Smith}}, \bibinfo {author}
  {\bibfnamefont {G.}~\bibnamefont {Vajente}}, \bibinfo {author} {\bibfnamefont
  {A.~R.}\ \bibnamefont {Williamson}}, \ and\ \bibinfo {author} {\bibfnamefont
  {C.~C.}\ \bibnamefont {Wipf}},\ }\href {\doibase
  10.1088/0264-9381/32/24/245005} {\bibfield  {journal} {\bibinfo  {journal}
  {Classical and Quantum Gravity}\ }\textbf {\bibinfo {volume} {32}},\ \bibinfo
  {pages} {245005} (\bibinfo {year} {2015})}\BibitemShut {NoStop}%
\bibitem [{\citenamefont {Messick}\ \emph {et~al.}(2017)\citenamefont
  {Messick}, \citenamefont {Blackburn}, \citenamefont {Brady}, \citenamefont
  {Brockill}, \citenamefont {Cannon}, \citenamefont {Cariou}, \citenamefont
  {Caudill}, \citenamefont {Chamberlin}, \citenamefont {Creighton},
  \citenamefont {Everett}, \citenamefont {Hanna}, \citenamefont {Keppel},
  \citenamefont {Lang}, \citenamefont {Li}, \citenamefont {Meacher},
  \citenamefont {Nielsen}, \citenamefont {Pankow}, \citenamefont {Privitera},
  \citenamefont {Qi}, \citenamefont {Sachdev}, \citenamefont {Sadeghian},
  \citenamefont {Singer}, \citenamefont {Thomas}, \citenamefont {Wade},
  \citenamefont {Wade}, \citenamefont {Weinstein},\ and\ \citenamefont
  {Wiesner}}]{messick2017analysis}%
  \BibitemOpen
  \bibfield  {author} {\bibinfo {author} {\bibfnamefont {C.}~\bibnamefont
  {Messick}}, \bibinfo {author} {\bibfnamefont {K.}~\bibnamefont {Blackburn}},
  \bibinfo {author} {\bibfnamefont {P.}~\bibnamefont {Brady}}, \bibinfo
  {author} {\bibfnamefont {P.}~\bibnamefont {Brockill}}, \bibinfo {author}
  {\bibfnamefont {K.}~\bibnamefont {Cannon}}, \bibinfo {author} {\bibfnamefont
  {R.}~\bibnamefont {Cariou}}, \bibinfo {author} {\bibfnamefont
  {S.}~\bibnamefont {Caudill}}, \bibinfo {author} {\bibfnamefont {S.~J.}\
  \bibnamefont {Chamberlin}}, \bibinfo {author} {\bibfnamefont {J.~D.~E.}\
  \bibnamefont {Creighton}}, \bibinfo {author} {\bibfnamefont {R.}~\bibnamefont
  {Everett}}, \bibinfo {author} {\bibfnamefont {C.}~\bibnamefont {Hanna}},
  \bibinfo {author} {\bibfnamefont {D.}~\bibnamefont {Keppel}}, \bibinfo
  {author} {\bibfnamefont {R.~N.}\ \bibnamefont {Lang}}, \bibinfo {author}
  {\bibfnamefont {T.~G.~F.}\ \bibnamefont {Li}}, \bibinfo {author}
  {\bibfnamefont {D.}~\bibnamefont {Meacher}}, \bibinfo {author} {\bibfnamefont
  {A.}~\bibnamefont {Nielsen}}, \bibinfo {author} {\bibfnamefont
  {C.}~\bibnamefont {Pankow}}, \bibinfo {author} {\bibfnamefont
  {S.}~\bibnamefont {Privitera}}, \bibinfo {author} {\bibfnamefont
  {H.}~\bibnamefont {Qi}}, \bibinfo {author} {\bibfnamefont {S.}~\bibnamefont
  {Sachdev}}, \bibinfo {author} {\bibfnamefont {L.}~\bibnamefont {Sadeghian}},
  \bibinfo {author} {\bibfnamefont {L.}~\bibnamefont {Singer}}, \bibinfo
  {author} {\bibfnamefont {E.~G.}\ \bibnamefont {Thomas}}, \bibinfo {author}
  {\bibfnamefont {L.}~\bibnamefont {Wade}}, \bibinfo {author} {\bibfnamefont
  {M.}~\bibnamefont {Wade}}, \bibinfo {author} {\bibfnamefont {A.}~\bibnamefont
  {Weinstein}}, \ and\ \bibinfo {author} {\bibfnamefont {K.}~\bibnamefont
  {Wiesner}},\ }\href {\doibase 10.1103/PhysRevD.95.042001} {\bibfield
  {journal} {\bibinfo  {journal} {Phys. Rev. D}\ }\textbf {\bibinfo {volume}
  {95}},\ \bibinfo {pages} {042001} (\bibinfo {year} {2017})}\BibitemShut
  {NoStop}%
\bibitem [{\citenamefont {Abbott}\ \emph
  {et~al.}(2016{\natexlab{e}})\citenamefont {Abbott} \emph
  {et~al.}}]{abbott2016binary}%
  \BibitemOpen
  \bibfield  {author} {\bibinfo {author} {\bibfnamefont {B.~P.}\ \bibnamefont
  {Abbott}} \emph {et~al.} (\bibinfo {collaboration} {LIGO Scientific
  Collaboration and Virgo Collaboration}),\ }\href {\doibase
  10.1103/PhysRevX.6.041015} {\bibfield  {journal} {\bibinfo  {journal} {Phys.
  Rev. X}\ }\textbf {\bibinfo {volume} {6}},\ \bibinfo {pages} {041015}
  (\bibinfo {year} {2016}{\natexlab{e}})}\BibitemShut {NoStop}%
\bibitem [{\citenamefont {Abbott}\ \emph
  {et~al.}(2017{\natexlab{e}})\citenamefont {Abbott} \emph
  {et~al.}}]{abbott2017search}%
  \BibitemOpen
  \bibfield  {author} {\bibinfo {author} {\bibfnamefont {B.~P.}\ \bibnamefont
  {Abbott}} \emph {et~al.} (\bibinfo {collaboration} {LIGO Scientific
  Collaboration and Virgo Collaboration}),\ }\href {\doibase
  10.1103/PhysRevD.96.022001} {\bibfield  {journal} {\bibinfo  {journal} {Phys.
  Rev. D}\ }\textbf {\bibinfo {volume} {96}},\ \bibinfo {pages} {022001}
  (\bibinfo {year} {2017}{\natexlab{e}})}\BibitemShut {NoStop}%
\bibitem [{\citenamefont {Abbott}\ \emph
  {et~al.}(2016{\natexlab{f}})\citenamefont {Abbott} \emph
  {et~al.}}]{abbott2016gw150914}%
  \BibitemOpen
  \bibfield  {author} {\bibinfo {author} {\bibfnamefont {B.~P.}\ \bibnamefont
  {Abbott}} \emph {et~al.} (\bibinfo {collaboration} {LIGO Scientific
  Collaboration and Virgo Collaboration}),\ }\href {\doibase
  10.1103/PhysRevD.93.122003} {\bibfield  {journal} {\bibinfo  {journal} {Phys.
  Rev. D}\ }\textbf {\bibinfo {volume} {93}},\ \bibinfo {pages} {122003}
  (\bibinfo {year} {2016}{\natexlab{f}})}\BibitemShut {NoStop}%
\bibitem [{\citenamefont {Abbott}\ \emph
  {et~al.}(2016{\natexlab{g}})\citenamefont {Abbott} \emph {et~al.}}]{BBH_O1}%
  \BibitemOpen
  \bibfield  {author} {\bibinfo {author} {\bibfnamefont {B.~P.}\ \bibnamefont
  {Abbott}} \emph {et~al.} (\bibinfo {collaboration} {LIGO Scientific
  Collaboration and Virgo Collaboration}),\ }\href {\doibase
  10.1103/PhysRevX.6.041015} {\bibfield  {journal} {\bibinfo  {journal} {Phys.
  Rev. X}\ }\textbf {\bibinfo {volume} {6}},\ \bibinfo {pages} {041015}
  (\bibinfo {year} {2016}{\natexlab{g}})}\BibitemShut {NoStop}%
\bibitem [{\citenamefont {Nitz}\ \emph {et~al.}(2020)\citenamefont {Nitz},
  \citenamefont {Dent}, \citenamefont {Davies}, \citenamefont {Kumar},
  \citenamefont {Capano} \emph {et~al.}}]{2OGC}%
  \BibitemOpen
  \bibfield  {author} {\bibinfo {author} {\bibfnamefont {A.~H.}\ \bibnamefont
  {Nitz}}, \bibinfo {author} {\bibfnamefont {T.}~\bibnamefont {Dent}}, \bibinfo
  {author} {\bibfnamefont {G.~S.}\ \bibnamefont {Davies}}, \bibinfo {author}
  {\bibfnamefont {S.}~\bibnamefont {Kumar}}, \bibinfo {author} {\bibfnamefont
  {C.~D.}\ \bibnamefont {Capano}},  \emph {et~al.},\ }\href {\doibase
  10.3847/1538-4357/ab733f} {\bibfield  {journal} {\bibinfo  {journal} {The
  Astrophysical Journal}\ }\textbf {\bibinfo {volume} {891}},\ \bibinfo {pages}
  {123} (\bibinfo {year} {2020})}\BibitemShut {NoStop}%
\bibitem [{\citenamefont {Szegedy}\ \emph {et~al.}(2016)\citenamefont
  {Szegedy}, \citenamefont {Vanhoucke}, \citenamefont {Ioffe}, \citenamefont
  {Shlens},\ and\ \citenamefont {Wojna}}]{szegedy2016rethinking}%
  \BibitemOpen
  \bibfield  {author} {\bibinfo {author} {\bibfnamefont {C.}~\bibnamefont
  {Szegedy}}, \bibinfo {author} {\bibfnamefont {V.}~\bibnamefont {Vanhoucke}},
  \bibinfo {author} {\bibfnamefont {S.}~\bibnamefont {Ioffe}}, \bibinfo
  {author} {\bibfnamefont {J.}~\bibnamefont {Shlens}}, \ and\ \bibinfo {author}
  {\bibfnamefont {Z.}~\bibnamefont {Wojna}},\ }in\ \href {\doibase
  10.1109/CVPR.2016.308} {\emph {\bibinfo {booktitle} {2016 IEEE Conference on
  Computer Vision and Pattern Recognition (CVPR)}}}\ (\bibinfo {year} {2016})\
  pp.\ \bibinfo {pages} {2818--2826}\BibitemShut {NoStop}%
\bibitem [{\citenamefont {Nitz}\ \emph {et~al.}(2017)\citenamefont {Nitz},
  \citenamefont {Dent}, \citenamefont {Canton}, \citenamefont {Fairhurst},\
  and\ \citenamefont {Brown}}]{improved_pycbc}%
  \BibitemOpen
  \bibfield  {author} {\bibinfo {author} {\bibfnamefont {A.~H.}\ \bibnamefont
  {Nitz}}, \bibinfo {author} {\bibfnamefont {T.}~\bibnamefont {Dent}}, \bibinfo
  {author} {\bibfnamefont {T.~D.}\ \bibnamefont {Canton}}, \bibinfo {author}
  {\bibfnamefont {S.}~\bibnamefont {Fairhurst}}, \ and\ \bibinfo {author}
  {\bibfnamefont {D.~A.}\ \bibnamefont {Brown}},\ }\href {\doibase
  10.3847/1538-4357/aa8f50} {\bibfield  {journal} {\bibinfo  {journal} {The
  Astrophysical Journal}\ }\textbf {\bibinfo {volume} {849}},\ \bibinfo {pages}
  {118} (\bibinfo {year} {2017})}\BibitemShut {NoStop}%
\bibitem [{\citenamefont {Venumadhav}\ \emph {et~al.}(2019)\citenamefont
  {Venumadhav}, \citenamefont {Zackay}, \citenamefont {Roulet}, \citenamefont
  {Dai},\ and\ \citenamefont {Zaldarriaga}}]{IAS_O1}%
  \BibitemOpen
  \bibfield  {author} {\bibinfo {author} {\bibfnamefont {T.}~\bibnamefont
  {Venumadhav}}, \bibinfo {author} {\bibfnamefont {B.}~\bibnamefont {Zackay}},
  \bibinfo {author} {\bibfnamefont {J.}~\bibnamefont {Roulet}}, \bibinfo
  {author} {\bibfnamefont {L.}~\bibnamefont {Dai}}, \ and\ \bibinfo {author}
  {\bibfnamefont {M.}~\bibnamefont {Zaldarriaga}},\ }\href {\doibase
  10.1103/PhysRevD.100.023011} {\bibfield  {journal} {\bibinfo  {journal}
  {Phys. Rev. D}\ }\textbf {\bibinfo {volume} {100}},\ \bibinfo {pages}
  {023011} (\bibinfo {year} {2019})}\BibitemShut {NoStop}%
\bibitem [{\citenamefont {Zackay}\ \emph {et~al.}(2019)\citenamefont {Zackay},
  \citenamefont {Venumadhav}, \citenamefont {Dai}, \citenamefont {Roulet},\
  and\ \citenamefont {Zaldarriaga}}]{IAS_GW151216}%
  \BibitemOpen
  \bibfield  {author} {\bibinfo {author} {\bibfnamefont {B.}~\bibnamefont
  {Zackay}}, \bibinfo {author} {\bibfnamefont {T.}~\bibnamefont {Venumadhav}},
  \bibinfo {author} {\bibfnamefont {L.}~\bibnamefont {Dai}}, \bibinfo {author}
  {\bibfnamefont {J.}~\bibnamefont {Roulet}}, \ and\ \bibinfo {author}
  {\bibfnamefont {M.}~\bibnamefont {Zaldarriaga}},\ }\href {\doibase
  10.1103/PhysRevD.100.023007} {\bibfield  {journal} {\bibinfo  {journal}
  {Phys. Rev. D}\ }\textbf {\bibinfo {volume} {100}},\ \bibinfo {pages}
  {023007} (\bibinfo {year} {2019})}\BibitemShut {NoStop}%
\bibitem [{\citenamefont {Gabbard}\ \emph {et~al.}(2018)\citenamefont
  {Gabbard}, \citenamefont {Williams}, \citenamefont {Hayes},\ and\
  \citenamefont {Messenger}}]{Matching_MF}%
  \BibitemOpen
  \bibfield  {author} {\bibinfo {author} {\bibfnamefont {H.}~\bibnamefont
  {Gabbard}}, \bibinfo {author} {\bibfnamefont {M.}~\bibnamefont {Williams}},
  \bibinfo {author} {\bibfnamefont {F.}~\bibnamefont {Hayes}}, \ and\ \bibinfo
  {author} {\bibfnamefont {C.}~\bibnamefont {Messenger}},\ }\href {\doibase
  10.1103/PhysRevLett.120.141103} {\bibfield  {journal} {\bibinfo  {journal}
  {Phys. Rev. Lett.}\ }\textbf {\bibinfo {volume} {120}},\ \bibinfo {pages}
  {141103} (\bibinfo {year} {2018})}\BibitemShut {NoStop}%
\bibitem [{\citenamefont {Sch\"afer}\ \emph {et~al.}(2020)\citenamefont
  {Sch\"afer}, \citenamefont {Ohme},\ and\ \citenamefont {Nitz}}]{BNS_ML}%
  \BibitemOpen
  \bibfield  {author} {\bibinfo {author} {\bibfnamefont {M.~B.}\ \bibnamefont
  {Sch\"afer}}, \bibinfo {author} {\bibfnamefont {F.}~\bibnamefont {Ohme}}, \
  and\ \bibinfo {author} {\bibfnamefont {A.~H.}\ \bibnamefont {Nitz}},\ }\href
  {\doibase 10.1103/PhysRevD.102.063015} {\bibfield  {journal} {\bibinfo
  {journal} {Phys. Rev. D}\ }\textbf {\bibinfo {volume} {102}},\ \bibinfo
  {pages} {063015} (\bibinfo {year} {2020})}\BibitemShut {NoStop}%
\bibitem [{\citenamefont {Cuoco}\ \emph {et~al.}(2020)\citenamefont {Cuoco},
  \citenamefont {Powell}, \citenamefont {Cavagli{\`{a}}}, \citenamefont
  {Ackley}, \citenamefont {Bejger} \emph {et~al.}}]{cuoco2020enhancing}%
  \BibitemOpen
  \bibfield  {author} {\bibinfo {author} {\bibfnamefont {E.}~\bibnamefont
  {Cuoco}}, \bibinfo {author} {\bibfnamefont {J.}~\bibnamefont {Powell}},
  \bibinfo {author} {\bibfnamefont {M.}~\bibnamefont {Cavagli{\`{a}}}},
  \bibinfo {author} {\bibfnamefont {K.}~\bibnamefont {Ackley}}, \bibinfo
  {author} {\bibfnamefont {M.}~\bibnamefont {Bejger}},  \emph {et~al.},\ }\href
  {\doibase 10.1088/2632-2153/abb93a} {\bibfield  {journal} {\bibinfo
  {journal} {Machine Learning: Science and Technology}\ }\textbf {\bibinfo
  {volume} {2}},\ \bibinfo {pages} {011002} (\bibinfo {year}
  {2020})}\BibitemShut {NoStop}%
\bibitem [{\citenamefont {Green}\ \emph {et~al.}(2020)\citenamefont {Green},
  \citenamefont {Simpson},\ and\ \citenamefont {Gair}}]{PE_AutoRegrFl}%
  \BibitemOpen
  \bibfield  {author} {\bibinfo {author} {\bibfnamefont {S.~R.}\ \bibnamefont
  {Green}}, \bibinfo {author} {\bibfnamefont {C.}~\bibnamefont {Simpson}}, \
  and\ \bibinfo {author} {\bibfnamefont {J.}~\bibnamefont {Gair}},\ }\href@noop
  {} {\enquote {\bibinfo {title} {Gravitational-wave parameter estimation with
  autoregressive neural network flows},}\ } (\bibinfo {year} {2020}),\ \Eprint
  {http://arxiv.org/abs/2002.07656} {arXiv:2002.07656 [astro-ph.IM]}
  \BibitemShut {NoStop}%
\bibitem [{\citenamefont {George}\ and\ \citenamefont
  {Huerta}(2018)}]{DNN_MMA}%
  \BibitemOpen
  \bibfield  {author} {\bibinfo {author} {\bibfnamefont {D.}~\bibnamefont
  {George}}\ and\ \bibinfo {author} {\bibfnamefont {E.~A.}\ \bibnamefont
  {Huerta}},\ }\href {\doibase 10.1103/PhysRevD.97.044039} {\bibfield
  {journal} {\bibinfo  {journal} {Phys. Rev. D}\ }\textbf {\bibinfo {volume}
  {97}},\ \bibinfo {pages} {044039} (\bibinfo {year} {2018})}\BibitemShut
  {NoStop}%
\bibitem [{\citenamefont {{George}}\ and\ \citenamefont
  {{Huerta}}(2018)}]{2018PhLB..778...64G}%
  \BibitemOpen
  \bibfield  {author} {\bibinfo {author} {\bibfnamefont {D.}~\bibnamefont
  {{George}}}\ and\ \bibinfo {author} {\bibfnamefont {E.~A.}\ \bibnamefont
  {{Huerta}}},\ }\href {\doibase 10.1016/j.physletb.2017.12.053} {\bibfield
  {journal} {\bibinfo  {journal} {PhLB}\ }\textbf {\bibinfo {volume} {778}},\
  \bibinfo {pages} {64} (\bibinfo {year} {2018})},\ \Eprint
  {http://arxiv.org/abs/1711.03121} {arXiv:1711.03121 [gr-qc]} \BibitemShut
  {NoStop}%
\bibitem [{\citenamefont {Chan}\ \emph {et~al.}(2020)\citenamefont {Chan},
  \citenamefont {Heng},\ and\ \citenamefont {Messenger}}]{Detecting_SN_DL}%
  \BibitemOpen
  \bibfield  {author} {\bibinfo {author} {\bibfnamefont {M.~L.}\ \bibnamefont
  {Chan}}, \bibinfo {author} {\bibfnamefont {I.~S.}\ \bibnamefont {Heng}}, \
  and\ \bibinfo {author} {\bibfnamefont {C.}~\bibnamefont {Messenger}},\ }\href
  {\doibase 10.1103/PhysRevD.102.043022} {\bibfield  {journal} {\bibinfo
  {journal} {Phys. Rev. D}\ }\textbf {\bibinfo {volume} {102}},\ \bibinfo
  {pages} {043022} (\bibinfo {year} {2020})}\BibitemShut {NoStop}%
\bibitem [{\citenamefont {Coughlin}\ \emph {et~al.}(2019)\citenamefont
  {Coughlin}, \citenamefont {Bahaadini}, \citenamefont {Rohani}, \citenamefont
  {Zevin}, \citenamefont {Patane}, \citenamefont {Harandi}, \citenamefont
  {Jackson}, \citenamefont {Noroozi}, \citenamefont {Allen}, \citenamefont
  {Areeda}, \citenamefont {Coughlin}, \citenamefont {Ruiz}, \citenamefont
  {Berry}, \citenamefont {Crowston}, \citenamefont {Katsaggelos}, \citenamefont
  {Lundgren}, \citenamefont {\O{}sterlund}, \citenamefont {Smith},
  \citenamefont {Trouille},\ and\ \citenamefont
  {Kalogera}}]{SimilarityLearning}%
  \BibitemOpen
  \bibfield  {author} {\bibinfo {author} {\bibfnamefont {S.}~\bibnamefont
  {Coughlin}}, \bibinfo {author} {\bibfnamefont {S.}~\bibnamefont {Bahaadini}},
  \bibinfo {author} {\bibfnamefont {N.}~\bibnamefont {Rohani}}, \bibinfo
  {author} {\bibfnamefont {M.}~\bibnamefont {Zevin}}, \bibinfo {author}
  {\bibfnamefont {O.}~\bibnamefont {Patane}}, \bibinfo {author} {\bibfnamefont
  {M.}~\bibnamefont {Harandi}}, \bibinfo {author} {\bibfnamefont
  {C.}~\bibnamefont {Jackson}}, \bibinfo {author} {\bibfnamefont
  {V.}~\bibnamefont {Noroozi}}, \bibinfo {author} {\bibfnamefont
  {S.}~\bibnamefont {Allen}}, \bibinfo {author} {\bibfnamefont
  {J.}~\bibnamefont {Areeda}}, \bibinfo {author} {\bibfnamefont
  {M.}~\bibnamefont {Coughlin}}, \bibinfo {author} {\bibfnamefont
  {P.}~\bibnamefont {Ruiz}}, \bibinfo {author} {\bibfnamefont {C.~P.~L.}\
  \bibnamefont {Berry}}, \bibinfo {author} {\bibfnamefont {K.}~\bibnamefont
  {Crowston}}, \bibinfo {author} {\bibfnamefont {A.~K.}\ \bibnamefont
  {Katsaggelos}}, \bibinfo {author} {\bibfnamefont {A.}~\bibnamefont
  {Lundgren}}, \bibinfo {author} {\bibfnamefont {C.}~\bibnamefont
  {\O{}sterlund}}, \bibinfo {author} {\bibfnamefont {J.~R.}\ \bibnamefont
  {Smith}}, \bibinfo {author} {\bibfnamefont {L.}~\bibnamefont {Trouille}}, \
  and\ \bibinfo {author} {\bibfnamefont {V.}~\bibnamefont {Kalogera}},\ }\href
  {\doibase 10.1103/PhysRevD.99.082002} {\bibfield  {journal} {\bibinfo
  {journal} {Phys. Rev. D}\ }\textbf {\bibinfo {volume} {99}},\ \bibinfo
  {pages} {082002} (\bibinfo {year} {2019})}\BibitemShut {NoStop}%
\bibitem [{\citenamefont {Marianer}\ \emph {et~al.}(2020)\citenamefont
  {Marianer}, \citenamefont {Poznanski},\ and\ \citenamefont
  {Prochaska}}]{10.1093/mnras/staa3550}%
  \BibitemOpen
  \bibfield  {author} {\bibinfo {author} {\bibfnamefont {T.}~\bibnamefont
  {Marianer}}, \bibinfo {author} {\bibfnamefont {D.}~\bibnamefont {Poznanski}},
  \ and\ \bibinfo {author} {\bibfnamefont {J.~X.}\ \bibnamefont {Prochaska}},\
  }\href {\doibase 10.1093/mnras/staa3550} {\bibfield  {journal} {\bibinfo
  {journal} {Monthly Notices of the Royal Astronomical Society}\ }\textbf
  {\bibinfo {volume} {500}},\ \bibinfo {pages} {5408} (\bibinfo {year}
  {2020})},\ \Eprint
  {http://arxiv.org/abs/https://academic.oup.com/mnras/article-pdf/500/4/5408/34925881/staa3550.pdf}
  {https://academic.oup.com/mnras/article-pdf/500/4/5408/34925881/staa3550.pdf}
  \BibitemShut {NoStop}%
\bibitem [{\citenamefont {Kim}\ \emph {et~al.}(2020)\citenamefont {Kim},
  \citenamefont {Li}, \citenamefont {Lo}, \citenamefont {Sachdev},\ and\
  \citenamefont {Yuen}}]{PhysRevD.101.083006}%
  \BibitemOpen
  \bibfield  {author} {\bibinfo {author} {\bibfnamefont {K.}~\bibnamefont
  {Kim}}, \bibinfo {author} {\bibfnamefont {T.~G.~F.}\ \bibnamefont {Li}},
  \bibinfo {author} {\bibfnamefont {R.~K.~L.}\ \bibnamefont {Lo}}, \bibinfo
  {author} {\bibfnamefont {S.}~\bibnamefont {Sachdev}}, \ and\ \bibinfo
  {author} {\bibfnamefont {R.~S.~H.}\ \bibnamefont {Yuen}},\ }\href {\doibase
  10.1103/PhysRevD.101.083006} {\bibfield  {journal} {\bibinfo  {journal}
  {Phys. Rev. D}\ }\textbf {\bibinfo {volume} {101}},\ \bibinfo {pages}
  {083006} (\bibinfo {year} {2020})}\BibitemShut {NoStop}%
\bibitem [{\citenamefont {{Miller}}\ \emph {et~al.}(2019)\citenamefont
  {{Miller}}, \citenamefont {{Astone}}, \citenamefont {{D'Antonio}},
  \citenamefont {{Frasca}}, \citenamefont {{Intini}}, \citenamefont {{La
  Rosa}}, \citenamefont {{Leaci}}, \citenamefont {{Mastrogiovanni}},
  \citenamefont {{Muciaccia}}, \citenamefont {{Mitidis}}, \citenamefont
  {{Palomba}}, \citenamefont {{Piccinni}}, \citenamefont {{Singhal}},
  \citenamefont {{Whiting}},\ and\ \citenamefont
  {{Rei}}}]{2019PhRvD.100f2005M}%
  \BibitemOpen
  \bibfield  {author} {\bibinfo {author} {\bibfnamefont {A.~L.}\ \bibnamefont
  {{Miller}}}, \bibinfo {author} {\bibfnamefont {P.}~\bibnamefont {{Astone}}},
  \bibinfo {author} {\bibfnamefont {S.}~\bibnamefont {{D'Antonio}}}, \bibinfo
  {author} {\bibfnamefont {S.}~\bibnamefont {{Frasca}}}, \bibinfo {author}
  {\bibfnamefont {G.}~\bibnamefont {{Intini}}}, \bibinfo {author}
  {\bibfnamefont {I.}~\bibnamefont {{La Rosa}}}, \bibinfo {author}
  {\bibfnamefont {P.}~\bibnamefont {{Leaci}}}, \bibinfo {author} {\bibfnamefont
  {S.}~\bibnamefont {{Mastrogiovanni}}}, \bibinfo {author} {\bibfnamefont
  {F.}~\bibnamefont {{Muciaccia}}}, \bibinfo {author} {\bibfnamefont
  {A.}~\bibnamefont {{Mitidis}}}, \bibinfo {author} {\bibfnamefont
  {C.}~\bibnamefont {{Palomba}}}, \bibinfo {author} {\bibfnamefont {O.~J.}\
  \bibnamefont {{Piccinni}}}, \bibinfo {author} {\bibfnamefont
  {A.}~\bibnamefont {{Singhal}}}, \bibinfo {author} {\bibfnamefont {B.~F.}\
  \bibnamefont {{Whiting}}}, \ and\ \bibinfo {author} {\bibfnamefont
  {L.}~\bibnamefont {{Rei}}},\ }\href {\doibase 10.1103/PhysRevD.100.062005}
  {\bibfield  {journal} {\bibinfo  {journal} {\prd}\ }\textbf {\bibinfo
  {volume} {100}},\ \bibinfo {eid} {062005} (\bibinfo {year} {2019})},\ \Eprint
  {http://arxiv.org/abs/1909.02262} {arXiv:1909.02262 [astro-ph.IM]}
  \BibitemShut {NoStop}%
\bibitem [{\citenamefont {{Gebhard}}\ \emph {et~al.}(2019)\citenamefont
  {{Gebhard}}, \citenamefont {{Kilbertus}}, \citenamefont {{Harry}},\ and\
  \citenamefont {{Sch{\"o}lkopf}}}]{2019PhRvD.100f3015G}%
  \BibitemOpen
  \bibfield  {author} {\bibinfo {author} {\bibfnamefont {T.~D.}\ \bibnamefont
  {{Gebhard}}}, \bibinfo {author} {\bibfnamefont {N.}~\bibnamefont
  {{Kilbertus}}}, \bibinfo {author} {\bibfnamefont {I.}~\bibnamefont
  {{Harry}}}, \ and\ \bibinfo {author} {\bibfnamefont {B.}~\bibnamefont
  {{Sch{\"o}lkopf}}},\ }\href {\doibase 10.1103/PhysRevD.100.063015} {\bibfield
   {journal} {\bibinfo  {journal} {\prd}\ }\textbf {\bibinfo {volume} {100}},\
  \bibinfo {eid} {063015} (\bibinfo {year} {2019})},\ \Eprint
  {http://arxiv.org/abs/1904.08693} {arXiv:1904.08693 [astro-ph.IM]}
  \BibitemShut {NoStop}%
\bibitem [{\citenamefont {Manteiga}\ \emph {et~al.}(2009)\citenamefont
  {Manteiga}, \citenamefont {Carricajo}, \citenamefont {Rodr{\'{\i}}guez},
  \citenamefont {Dafonte},\ and\ \citenamefont {Arcay}}]{Manteiga_2009}%
  \BibitemOpen
  \bibfield  {author} {\bibinfo {author} {\bibfnamefont {M.}~\bibnamefont
  {Manteiga}}, \bibinfo {author} {\bibfnamefont {I.}~\bibnamefont {Carricajo}},
  \bibinfo {author} {\bibfnamefont {A.}~\bibnamefont {Rodr{\'{\i}}guez}},
  \bibinfo {author} {\bibfnamefont {C.}~\bibnamefont {Dafonte}}, \ and\
  \bibinfo {author} {\bibfnamefont {B.}~\bibnamefont {Arcay}},\ }\href
  {\doibase 10.1088/0004-6256/137/2/3245} {\bibfield  {journal} {\bibinfo
  {journal} {The Astronomical Journal}\ }\textbf {\bibinfo {volume} {137}},\
  \bibinfo {pages} {3245} (\bibinfo {year} {2009})}\BibitemShut {NoStop}%
\bibitem [{\citenamefont {Almeida}\ and\ \citenamefont
  {Prieto}(2013)}]{S_nchez_Almeida_2013}%
  \BibitemOpen
  \bibfield  {author} {\bibinfo {author} {\bibfnamefont {J.~S.}\ \bibnamefont
  {Almeida}}\ and\ \bibinfo {author} {\bibfnamefont {C.~A.}\ \bibnamefont
  {Prieto}},\ }\href {\doibase 10.1088/0004-637x/763/1/50} {\bibfield
  {journal} {\bibinfo  {journal} {The Astrophysical Journal}\ }\textbf
  {\bibinfo {volume} {763}},\ \bibinfo {pages} {50} (\bibinfo {year}
  {2013})}\BibitemShut {NoStop}%
\bibitem [{\citenamefont {{Kuntzer}}\ \emph {et~al.}(2016)\citenamefont
  {{Kuntzer}}, \citenamefont {{Tewes}},\ and\ \citenamefont
  {{Courbin}}}]{kuntzer}%
  \BibitemOpen
  \bibfield  {author} {\bibinfo {author} {\bibfnamefont {T.}~\bibnamefont
  {{Kuntzer}}}, \bibinfo {author} {\bibfnamefont {M.}~\bibnamefont {{Tewes}}},
  \ and\ \bibinfo {author} {\bibfnamefont {F.}~\bibnamefont {{Courbin}}},\
  }\href {\doibase 10.1051/0004-6361/201628660} {\bibfield  {journal} {\bibinfo
   {journal} {Astron. \& Astrophys.}\ }\textbf {\bibinfo {volume} {591}},\
  \bibinfo {eid} {A54} (\bibinfo {year} {2016})},\ \Eprint
  {http://arxiv.org/abs/1605.03201} {arXiv:1605.03201 [astro-ph.IM]}
  \BibitemShut {NoStop}%
\bibitem [{\citenamefont {Abraham}\ \emph {et~al.}(2020)\citenamefont
  {Abraham}, \citenamefont {Mukund}, \citenamefont {Vibhute}, \citenamefont
  {Sharma}, \citenamefont {Iyyani} \emph {et~al.}}]{abraham2020machine}%
  \BibitemOpen
  \bibfield  {author} {\bibinfo {author} {\bibfnamefont {S.}~\bibnamefont
  {Abraham}}, \bibinfo {author} {\bibfnamefont {N.}~\bibnamefont {Mukund}},
  \bibinfo {author} {\bibfnamefont {A.}~\bibnamefont {Vibhute}}, \bibinfo
  {author} {\bibfnamefont {V.}~\bibnamefont {Sharma}}, \bibinfo {author}
  {\bibfnamefont {S.}~\bibnamefont {Iyyani}},  \emph {et~al.},\ }\href@noop {}
  {\enquote {\bibinfo {title} {A machine learning approach for grb detection in
  astrosat czti data},}\ } (\bibinfo {year} {2020}),\ \Eprint
  {http://arxiv.org/abs/1906.09670} {arXiv:1906.09670 [astro-ph.IM]}
  \BibitemShut {NoStop}%
\bibitem [{\citenamefont {{Philip}}\ \emph {et~al.}(2002)\citenamefont
  {{Philip}}, \citenamefont {{Wadadekar}}, \citenamefont {{Kembhavi}},\ and\
  \citenamefont {{Joseph}}}]{star_galaxy_dbnn}%
  \BibitemOpen
  \bibfield  {author} {\bibinfo {author} {\bibfnamefont {N.~S.}\ \bibnamefont
  {{Philip}}}, \bibinfo {author} {\bibfnamefont {Y.}~\bibnamefont
  {{Wadadekar}}}, \bibinfo {author} {\bibfnamefont {A.}~\bibnamefont
  {{Kembhavi}}}, \ and\ \bibinfo {author} {\bibfnamefont {K.~B.}\ \bibnamefont
  {{Joseph}}},\ }\href {\doibase 10.1051/0004-6361:20020219} {\bibfield
  {journal} {\bibinfo  {journal} {Astron. \& Astrophys.}\ }\textbf {\bibinfo
  {volume} {385}},\ \bibinfo {pages} {1119} (\bibinfo {year} {2002})},\ \Eprint
  {http://arxiv.org/abs/astro-ph/0202127} {arXiv:astro-ph/0202127 [astro-ph]}
  \BibitemShut {NoStop}%
\bibitem [{\citenamefont {{Odewahn}}\ \emph {et~al.}(2002)\citenamefont
  {{Odewahn}}, \citenamefont {{Cohen}}, \citenamefont {{Windhorst}},\ and\
  \citenamefont {{Philip}}}]{auto_galax_morph}%
  \BibitemOpen
  \bibfield  {author} {\bibinfo {author} {\bibfnamefont {S.~C.}\ \bibnamefont
  {{Odewahn}}}, \bibinfo {author} {\bibfnamefont {S.~H.}\ \bibnamefont
  {{Cohen}}}, \bibinfo {author} {\bibfnamefont {R.~A.}\ \bibnamefont
  {{Windhorst}}}, \ and\ \bibinfo {author} {\bibfnamefont {N.~S.}\ \bibnamefont
  {{Philip}}},\ }\href {\doibase 10.1086/339036} {\bibfield  {journal}
  {\bibinfo  {journal} {Astrophys. J.}\ }\textbf {\bibinfo {volume} {568}},\
  \bibinfo {pages} {539} (\bibinfo {year} {2002})},\ \Eprint
  {http://arxiv.org/abs/astro-ph/0110275} {arXiv:astro-ph/0110275 [astro-ph]}
  \BibitemShut {NoStop}%
\bibitem [{\citenamefont {Li}\ \emph {et~al.}(2019)\citenamefont {Li},
  \citenamefont {Lin},\ and\ \citenamefont {Qiu}}]{randforest}%
  \BibitemOpen
  \bibfield  {author} {\bibinfo {author} {\bibfnamefont {X.-R.}\ \bibnamefont
  {Li}}, \bibinfo {author} {\bibfnamefont {Y.-T.}\ \bibnamefont {Lin}}, \ and\
  \bibinfo {author} {\bibfnamefont {K.-B.}\ \bibnamefont {Qiu}},\ }\href
  {\doibase 10.1088/1674-4527/19/8/111} {\bibfield  {journal} {\bibinfo
  {journal} {Research in Astronomy and Astrophysics}\ }\textbf {\bibinfo
  {volume} {19}},\ \bibinfo {pages} {111} (\bibinfo {year} {2019})}\BibitemShut
  {NoStop}%
\bibitem [{\citenamefont {Richards}\ \emph {et~al.}(2009)\citenamefont
  {Richards}, \citenamefont {Myers}, \citenamefont {Gray}, \citenamefont
  {Riegel}, \citenamefont {Nichol} \emph {et~al.}}]{Richards_2008}%
  \BibitemOpen
  \bibfield  {author} {\bibinfo {author} {\bibfnamefont {G.~T.}\ \bibnamefont
  {Richards}}, \bibinfo {author} {\bibfnamefont {A.~D.}\ \bibnamefont {Myers}},
  \bibinfo {author} {\bibfnamefont {A.~G.}\ \bibnamefont {Gray}}, \bibinfo
  {author} {\bibfnamefont {R.~N.}\ \bibnamefont {Riegel}}, \bibinfo {author}
  {\bibfnamefont {R.~C.}\ \bibnamefont {Nichol}},  \emph {et~al.},\ }\href
  {\doibase 10.1088/0067-0049/180/1/67} {\bibfield  {journal} {\bibinfo
  {journal} {The Astrophysical Journal Supplement Series}\ }\textbf {\bibinfo
  {volume} {180}},\ \bibinfo {pages} {67} (\bibinfo {year} {2009})}\BibitemShut
  {NoStop}%
\bibitem [{\citenamefont {{Abraham}}\ \emph {et~al.}(2012)\citenamefont
  {{Abraham}}, \citenamefont {{Philip}}, \citenamefont {{Kembhavi}},
  \citenamefont {{Wadadekar}},\ and\ \citenamefont {{Sinha}}}]{abraham_sdss}%
  \BibitemOpen
  \bibfield  {author} {\bibinfo {author} {\bibfnamefont {S.}~\bibnamefont
  {{Abraham}}}, \bibinfo {author} {\bibfnamefont {N.~S.}\ \bibnamefont
  {{Philip}}}, \bibinfo {author} {\bibfnamefont {A.}~\bibnamefont
  {{Kembhavi}}}, \bibinfo {author} {\bibfnamefont {Y.~G.}\ \bibnamefont
  {{Wadadekar}}}, \ and\ \bibinfo {author} {\bibfnamefont {R.}~\bibnamefont
  {{Sinha}}},\ }\href {\doibase 10.1111/j.1365-2966.2011.19674.x} {\bibfield
  {journal} {\bibinfo  {journal} {Mon. Not. R. Astron. Soc.}\ }\textbf
  {\bibinfo {volume} {419}},\ \bibinfo {pages} {80} (\bibinfo {year}
  {2012})}\BibitemShut {NoStop}%
\bibitem [{\citenamefont {Peters}\ \emph {et~al.}(2015)\citenamefont {Peters}
  \emph {et~al.}}]{Peters_2015}%
  \BibitemOpen
  \bibfield  {author} {\bibinfo {author} {\bibfnamefont {C.~M.}\ \bibnamefont
  {Peters}} \emph {et~al.},\ }\href {\doibase 10.1088/0004-637x/811/2/95}
  {\bibfield  {journal} {\bibinfo  {journal} {The Astrophysical Journal}\
  }\textbf {\bibinfo {volume} {811}},\ \bibinfo {pages} {95} (\bibinfo {year}
  {2015})}\BibitemShut {NoStop}%
\bibitem [{\citenamefont {Mukund}\ \emph {et~al.}(2017)\citenamefont {Mukund},
  \citenamefont {Abraham}, \citenamefont {Kandhasamy}, \citenamefont {Mitra},\
  and\ \citenamefont {Philip}}]{dbnn}%
  \BibitemOpen
  \bibfield  {author} {\bibinfo {author} {\bibfnamefont {N.}~\bibnamefont
  {Mukund}}, \bibinfo {author} {\bibfnamefont {S.}~\bibnamefont {Abraham}},
  \bibinfo {author} {\bibfnamefont {S.}~\bibnamefont {Kandhasamy}}, \bibinfo
  {author} {\bibfnamefont {S.}~\bibnamefont {Mitra}}, \ and\ \bibinfo {author}
  {\bibfnamefont {N.~S.}\ \bibnamefont {Philip}},\ }\href {\doibase
  10.1103/PhysRevD.95.104059} {\bibfield  {journal} {\bibinfo  {journal} {Phys.
  Rev. D}\ }\textbf {\bibinfo {volume} {95}},\ \bibinfo {pages} {104059}
  (\bibinfo {year} {2017})}\BibitemShut {NoStop}%
\bibitem [{\citenamefont {Zevin}\ \emph {et~al.}(2017)\citenamefont {Zevin},
  \citenamefont {Coughlin}, \citenamefont {Bahaadini}, \citenamefont {Besler},
  \citenamefont {Rohani} \emph {et~al.}}]{gravityspy}%
  \BibitemOpen
  \bibfield  {author} {\bibinfo {author} {\bibfnamefont {M.}~\bibnamefont
  {Zevin}}, \bibinfo {author} {\bibfnamefont {S.}~\bibnamefont {Coughlin}},
  \bibinfo {author} {\bibfnamefont {S.}~\bibnamefont {Bahaadini}}, \bibinfo
  {author} {\bibfnamefont {E.}~\bibnamefont {Besler}}, \bibinfo {author}
  {\bibfnamefont {N.}~\bibnamefont {Rohani}},  \emph {et~al.},\ }\href
  {\doibase 10.1088/1361-6382/aa5cea} {\bibfield  {journal} {\bibinfo
  {journal} {Classical and Quantum Gravity}\ }\textbf {\bibinfo {volume}
  {34}},\ \bibinfo {pages} {064003} (\bibinfo {year} {2017})}\BibitemShut
  {NoStop}%
\bibitem [{\citenamefont {Deng}\ \emph {et~al.}(2009)\citenamefont {Deng},
  \citenamefont {Dong}, \citenamefont {Socher}, \citenamefont {Li},
  \citenamefont {Li},\ and\ \citenamefont {Fei-Fei}}]{deng2009imagenet}%
  \BibitemOpen
  \bibfield  {author} {\bibinfo {author} {\bibfnamefont {J.}~\bibnamefont
  {Deng}}, \bibinfo {author} {\bibfnamefont {W.}~\bibnamefont {Dong}}, \bibinfo
  {author} {\bibfnamefont {R.}~\bibnamefont {Socher}}, \bibinfo {author}
  {\bibfnamefont {L.-J.}\ \bibnamefont {Li}}, \bibinfo {author} {\bibfnamefont
  {K.}~\bibnamefont {Li}}, \ and\ \bibinfo {author} {\bibfnamefont
  {L.}~\bibnamefont {Fei-Fei}},\ }in\ \href@noop {} {\emph {\bibinfo
  {booktitle} {2009 IEEE conference on computer vision and pattern
  recognition}}}\ (\bibinfo {organization} {Ieee},\ \bibinfo {year} {2009})\
  pp.\ \bibinfo {pages} {248--255}\BibitemShut {NoStop}%
\bibitem [{\citenamefont {Krizhevsky}\ \emph {et~al.}(2012)\citenamefont
  {Krizhevsky}, \citenamefont {Sutskever},\ and\ \citenamefont
  {Hinton}}]{krizhevsky2012imagenet}%
  \BibitemOpen
  \bibfield  {author} {\bibinfo {author} {\bibfnamefont {A.}~\bibnamefont
  {Krizhevsky}}, \bibinfo {author} {\bibfnamefont {I.}~\bibnamefont
  {Sutskever}}, \ and\ \bibinfo {author} {\bibfnamefont {G.~E.}\ \bibnamefont
  {Hinton}},\ }in\ \href@noop {} {\emph {\bibinfo {booktitle} {Advances in
  neural information processing systems}}}\ (\bibinfo {year} {2012})\ pp.\
  \bibinfo {pages} {1097--1105}\BibitemShut {NoStop}%
\bibitem [{\citenamefont {He}\ \emph {et~al.}(2016)\citenamefont {He},
  \citenamefont {Zhang}, \citenamefont {Ren},\ and\ \citenamefont
  {Sun}}]{resnet}%
  \BibitemOpen
  \bibfield  {author} {\bibinfo {author} {\bibfnamefont {K.}~\bibnamefont
  {He}}, \bibinfo {author} {\bibfnamefont {X.}~\bibnamefont {Zhang}}, \bibinfo
  {author} {\bibfnamefont {S.}~\bibnamefont {Ren}}, \ and\ \bibinfo {author}
  {\bibfnamefont {J.}~\bibnamefont {Sun}},\ }in\ \href@noop {} {\emph {\bibinfo
  {booktitle} {Proceedings of the IEEE conference on computer vision and
  pattern recognition}}}\ (\bibinfo {year} {2016})\ pp.\ \bibinfo {pages}
  {770--778}\BibitemShut {NoStop}%
\bibitem [{\citenamefont {Szegedy}\ \emph {et~al.}(2015)\citenamefont
  {Szegedy}, \citenamefont {Liu}, \citenamefont {Jia}, \citenamefont
  {Sermanet}, \citenamefont {Reed} \emph {et~al.}}]{szegedy2015going}%
  \BibitemOpen
  \bibfield  {author} {\bibinfo {author} {\bibfnamefont {C.}~\bibnamefont
  {Szegedy}}, \bibinfo {author} {\bibfnamefont {W.}~\bibnamefont {Liu}},
  \bibinfo {author} {\bibfnamefont {Y.}~\bibnamefont {Jia}}, \bibinfo {author}
  {\bibfnamefont {P.}~\bibnamefont {Sermanet}}, \bibinfo {author}
  {\bibfnamefont {S.}~\bibnamefont {Reed}},  \emph {et~al.},\ }in\ \href@noop
  {} {\emph {\bibinfo {booktitle} {Proceedings of the IEEE conference on
  computer vision and pattern recognition}}}\ (\bibinfo {year} {2015})\ pp.\
  \bibinfo {pages} {1--9}\BibitemShut {NoStop}%
\bibitem [{\citenamefont {Creighton}\ and\ \citenamefont
  {Anderson}(2012)}]{creighton2012gravitational}%
  \BibitemOpen
  \bibfield  {author} {\bibinfo {author} {\bibfnamefont {J.~D.}\ \bibnamefont
  {Creighton}}\ and\ \bibinfo {author} {\bibfnamefont {W.~G.}\ \bibnamefont
  {Anderson}},\ }\href@noop {} {\emph {\bibinfo {title} {Gravitational-wave
  physics and astronomy: An introduction to theory, experiment and data
  analysis}}}\ (\bibinfo  {publisher} {John Wiley \& Sons},\ \bibinfo {year}
  {2012})\BibitemShut {NoStop}%
\bibitem [{\citenamefont {Debnath}\ and\ \citenamefont
  {Shah}(2002)}]{debnath2002wavelet}%
  \BibitemOpen
  \bibfield  {author} {\bibinfo {author} {\bibfnamefont {L.}~\bibnamefont
  {Debnath}}\ and\ \bibinfo {author} {\bibfnamefont {F.~A.}\ \bibnamefont
  {Shah}},\ }\href@noop {} {\emph {\bibinfo {title} {Wavelet transforms and
  their applications}}}\ (\bibinfo  {publisher} {Springer},\ \bibinfo {year}
  {2002})\BibitemShut {NoStop}%
\bibitem [{\citenamefont {Bahaadini}\ \emph {et~al.}(2018)\citenamefont
  {Bahaadini}, \citenamefont {Noroozi}, \citenamefont {Rohani}, \citenamefont
  {Coughlin}, \citenamefont {Zevin}, \citenamefont {Smith}, \citenamefont
  {Kalogera},\ and\ \citenamefont {Katsaggelos}}]{GravitySpyDb}%
  \BibitemOpen
  \bibfield  {author} {\bibinfo {author} {\bibfnamefont {S.}~\bibnamefont
  {Bahaadini}}, \bibinfo {author} {\bibfnamefont {V.}~\bibnamefont {Noroozi}},
  \bibinfo {author} {\bibfnamefont {N.}~\bibnamefont {Rohani}}, \bibinfo
  {author} {\bibfnamefont {S.}~\bibnamefont {Coughlin}}, \bibinfo {author}
  {\bibfnamefont {M.}~\bibnamefont {Zevin}}, \bibinfo {author} {\bibfnamefont
  {J.}~\bibnamefont {Smith}}, \bibinfo {author} {\bibfnamefont
  {V.}~\bibnamefont {Kalogera}}, \ and\ \bibinfo {author} {\bibfnamefont
  {A.}~\bibnamefont {Katsaggelos}},\ }\href {\doibase
  https://doi.org/10.1016/j.ins.2018.02.068} {\bibfield  {journal} {\bibinfo
  {journal} {Information Sciences}\ }\textbf {\bibinfo {volume} {444}},\
  \bibinfo {pages} {172 } (\bibinfo {year} {2018})}\BibitemShut {NoStop}%
\bibitem [{\citenamefont {Abbott}\ \emph {et~al.}(2009)\citenamefont {Abbott}
  \emph {et~al.}}]{PhysRevD.79.122001}%
  \BibitemOpen
  \bibfield  {author} {\bibinfo {author} {\bibfnamefont {B.~P.}\ \bibnamefont
  {Abbott}} \emph {et~al.} (\bibinfo {collaboration} {LIGO Scientific
  Collaboration}),\ }\href {\doibase 10.1103/PhysRevD.79.122001} {\bibfield
  {journal} {\bibinfo  {journal} {Phys. Rev. D}\ }\textbf {\bibinfo {volume}
  {79}},\ \bibinfo {pages} {122001} (\bibinfo {year} {2009})}\BibitemShut
  {NoStop}%
\bibitem [{\citenamefont {Boh\'e}\ \emph {et~al.}(2017)\citenamefont {Boh\'e},
  \citenamefont {Shao}, \citenamefont {Taracchini}, \citenamefont {Buonanno},
  \citenamefont {Babak}, \citenamefont {Harry}, \citenamefont {Hinder},
  \citenamefont {Ossokine}, \citenamefont {P\"urrer}, \citenamefont {Raymond},
  \citenamefont {Chu}, \citenamefont {Fong}, \citenamefont {Kumar},
  \citenamefont {Pfeiffer}, \citenamefont {Boyle}, \citenamefont {Hemberger},
  \citenamefont {Kidder}, \citenamefont {Lovelace}, \citenamefont {Scheel},\
  and\ \citenamefont {Szil\'agyi}}]{seobnrv4}%
  \BibitemOpen
  \bibfield  {author} {\bibinfo {author} {\bibfnamefont {A.}~\bibnamefont
  {Boh\'e}}, \bibinfo {author} {\bibfnamefont {L.}~\bibnamefont {Shao}},
  \bibinfo {author} {\bibfnamefont {A.}~\bibnamefont {Taracchini}}, \bibinfo
  {author} {\bibfnamefont {A.}~\bibnamefont {Buonanno}}, \bibinfo {author}
  {\bibfnamefont {S.}~\bibnamefont {Babak}}, \bibinfo {author} {\bibfnamefont
  {I.~W.}\ \bibnamefont {Harry}}, \bibinfo {author} {\bibfnamefont
  {I.}~\bibnamefont {Hinder}}, \bibinfo {author} {\bibfnamefont
  {S.}~\bibnamefont {Ossokine}}, \bibinfo {author} {\bibfnamefont
  {M.}~\bibnamefont {P\"urrer}}, \bibinfo {author} {\bibfnamefont
  {V.}~\bibnamefont {Raymond}}, \bibinfo {author} {\bibfnamefont
  {T.}~\bibnamefont {Chu}}, \bibinfo {author} {\bibfnamefont {H.}~\bibnamefont
  {Fong}}, \bibinfo {author} {\bibfnamefont {P.}~\bibnamefont {Kumar}},
  \bibinfo {author} {\bibfnamefont {H.~P.}\ \bibnamefont {Pfeiffer}}, \bibinfo
  {author} {\bibfnamefont {M.}~\bibnamefont {Boyle}}, \bibinfo {author}
  {\bibfnamefont {D.~A.}\ \bibnamefont {Hemberger}}, \bibinfo {author}
  {\bibfnamefont {L.~E.}\ \bibnamefont {Kidder}}, \bibinfo {author}
  {\bibfnamefont {G.}~\bibnamefont {Lovelace}}, \bibinfo {author}
  {\bibfnamefont {M.~A.}\ \bibnamefont {Scheel}}, \ and\ \bibinfo {author}
  {\bibfnamefont {B.}~\bibnamefont {Szil\'agyi}},\ }\href {\doibase
  10.1103/PhysRevD.95.044028} {\bibfield  {journal} {\bibinfo  {journal} {Phys.
  Rev. D}\ }\textbf {\bibinfo {volume} {95}},\ \bibinfo {pages} {044028}
  (\bibinfo {year} {2017})}\BibitemShut {NoStop}%
\bibitem [{\citenamefont {Qian}(1999)}]{qian1999momentum}%
  \BibitemOpen
  \bibfield  {author} {\bibinfo {author} {\bibfnamefont {N.}~\bibnamefont
  {Qian}},\ }\href {\doibase https://doi.org/10.1016/S0893-6080(98)00116-6}
  {\bibfield  {journal} {\bibinfo  {journal} {Neural Networks}\ }\textbf
  {\bibinfo {volume} {12}},\ \bibinfo {pages} {145 } (\bibinfo {year}
  {1999})}\BibitemShut {NoStop}%
\bibitem [{\citenamefont {Allen}(2005)}]{Allen_chi}%
  \BibitemOpen
  \bibfield  {author} {\bibinfo {author} {\bibfnamefont {B.}~\bibnamefont
  {Allen}},\ }\href {\doibase 10.1103/PhysRevD.71.062001} {\bibfield  {journal}
  {\bibinfo  {journal} {Phys. Rev. D}\ }\textbf {\bibinfo {volume} {71}},\
  \bibinfo {pages} {062001} (\bibinfo {year} {2005})}\BibitemShut {NoStop}%
\bibitem [{\citenamefont {Nitz}(2018)}]{sgveto}%
  \BibitemOpen
  \bibfield  {author} {\bibinfo {author} {\bibfnamefont {A.~H.}\ \bibnamefont
  {Nitz}},\ }\href {\doibase 10.1088/1361-6382/aaa13d} {\bibfield  {journal}
  {\bibinfo  {journal} {Classical and Quantum Gravity}\ }\textbf {\bibinfo
  {volume} {35}},\ \bibinfo {pages} {035016} (\bibinfo {year}
  {2018})}\BibitemShut {NoStop}%
\bibitem [{\citenamefont {Ajith}(2011)}]{SpinTaylorT5}%
  \BibitemOpen
  \bibfield  {author} {\bibinfo {author} {\bibfnamefont {P.}~\bibnamefont
  {Ajith}},\ }\href {\doibase 10.1103/PhysRevD.84.084037} {\bibfield  {journal}
  {\bibinfo  {journal} {Phys. Rev. D}\ }\textbf {\bibinfo {volume} {84}},\
  \bibinfo {pages} {084037} (\bibinfo {year} {2011})}\BibitemShut {NoStop}%
\bibitem [{git()}]{gitlink}%
  \BibitemOpen
  \href@noop {} {\enquote {\bibinfo {title} {{MLStat search data release}},}\
  }\bibinfo {howpublished}
  {\url{https://github.com/shreejit-92/MLStat-1.git}}\BibitemShut {NoStop}%
\bibitem [{\citenamefont {Canton}\ and\ \citenamefont
  {Harry}(2017)}]{canton2017designing}%
  \BibitemOpen
  \bibfield  {author} {\bibinfo {author} {\bibfnamefont {T.~D.}\ \bibnamefont
  {Canton}}\ and\ \bibinfo {author} {\bibfnamefont {I.~W.}\ \bibnamefont
  {Harry}},\ }\href@noop {} {\enquote {\bibinfo {title} {Designing a template
  bank to observe compact binary coalescences in advanced ligo's second
  observing run},}\ } (\bibinfo {year} {2017}),\ \Eprint
  {http://arxiv.org/abs/1705.01845} {arXiv:1705.01845 [gr-qc]} \BibitemShut
  {NoStop}%
\bibitem [{\citenamefont {Ashton}\ and\ \citenamefont
  {Thrane}(2020)}]{odds_GW151216}%
  \BibitemOpen
  \bibfield  {author} {\bibinfo {author} {\bibfnamefont {G.}~\bibnamefont
  {Ashton}}\ and\ \bibinfo {author} {\bibfnamefont {E.}~\bibnamefont
  {Thrane}},\ }\href {\doibase 10.1093/mnras/staa2332} {\bibfield  {journal}
  {\bibinfo  {journal} {Monthly Notices of the Royal Astronomical Society}\
  }\textbf {\bibinfo {volume} {498}},\ \bibinfo {pages} {1905} (\bibinfo {year}
  {2020})},\ \Eprint
  {http://arxiv.org/abs/https://academic.oup.com/mnras/article-pdf/498/2/1905/33752535/staa2332.pdf}
  {https://academic.oup.com/mnras/article-pdf/498/2/1905/33752535/staa2332.pdf}
  \BibitemShut {NoStop}%
\bibitem [{\citenamefont {Pratten}\ and\ \citenamefont
  {Vecchio}(2020)}]{pratten2020assessing}%
  \BibitemOpen
  \bibfield  {author} {\bibinfo {author} {\bibfnamefont {G.}~\bibnamefont
  {Pratten}}\ and\ \bibinfo {author} {\bibfnamefont {A.}~\bibnamefont
  {Vecchio}},\ }\href@noop {} {\enquote {\bibinfo {title} {Assessing
  gravitational-wave binary black hole candidates with bayesian odds},}\ }
  (\bibinfo {year} {2020}),\ \Eprint {http://arxiv.org/abs/2008.00509}
  {arXiv:2008.00509 [gr-qc]} \BibitemShut {NoStop}%
\bibitem [{\citenamefont {Huang}\ \emph {et~al.}(2020)\citenamefont {Huang},
  \citenamefont {Haster}, \citenamefont {Vitale}, \citenamefont {Zimmerman},
  \citenamefont {Roulet}, \citenamefont {Venumadhav}, \citenamefont {Zackay},
  \citenamefont {Dai},\ and\ \citenamefont {Zaldarriaga}}]{huang2020source}%
  \BibitemOpen
  \bibfield  {author} {\bibinfo {author} {\bibfnamefont {Y.}~\bibnamefont
  {Huang}}, \bibinfo {author} {\bibfnamefont {C.-J.}\ \bibnamefont {Haster}},
  \bibinfo {author} {\bibfnamefont {S.}~\bibnamefont {Vitale}}, \bibinfo
  {author} {\bibfnamefont {A.}~\bibnamefont {Zimmerman}}, \bibinfo {author}
  {\bibfnamefont {J.}~\bibnamefont {Roulet}}, \bibinfo {author} {\bibfnamefont
  {T.}~\bibnamefont {Venumadhav}}, \bibinfo {author} {\bibfnamefont
  {B.}~\bibnamefont {Zackay}}, \bibinfo {author} {\bibfnamefont
  {L.}~\bibnamefont {Dai}}, \ and\ \bibinfo {author} {\bibfnamefont
  {M.}~\bibnamefont {Zaldarriaga}},\ }\href@noop {} {\enquote {\bibinfo {title}
  {Source properties of the lowest signal-to-noise-ratio binary black hole
  detections},}\ } (\bibinfo {year} {2020}),\ \Eprint
  {http://arxiv.org/abs/2003.04513} {arXiv:2003.04513 [gr-qc]} \BibitemShut
  {NoStop}%
\bibitem [{\citenamefont {Ashton}\ \emph {et~al.}(2019)\citenamefont {Ashton}
  \emph {et~al.}}]{Ashton:2018jfp}%
  \BibitemOpen
  \bibfield  {author} {\bibinfo {author} {\bibfnamefont {G.}~\bibnamefont
  {Ashton}} \emph {et~al.},\ }\href {\doibase 10.3847/1538-4365/ab06fc}
  {\bibfield  {journal} {\bibinfo  {journal} {Astrophys. J. Suppl.}\ }\textbf
  {\bibinfo {volume} {241}},\ \bibinfo {pages} {27} (\bibinfo {year} {2019})},\
  \Eprint {http://arxiv.org/abs/1811.02042} {arXiv:1811.02042 [astro-ph.IM]}
  \BibitemShut {NoStop}%
\bibitem [{\citenamefont {Romero-Shaw}\ \emph {et~al.}(2020)\citenamefont
  {Romero-Shaw}, \citenamefont {Talbot}, \citenamefont {Biscoveanu},
  \citenamefont {D’Emilio}, \citenamefont {Ashton} \emph
  {et~al.}}]{bilby_gwtc1}%
  \BibitemOpen
  \bibfield  {author} {\bibinfo {author} {\bibfnamefont {I.~M.}\ \bibnamefont
  {Romero-Shaw}}, \bibinfo {author} {\bibfnamefont {C.}~\bibnamefont {Talbot}},
  \bibinfo {author} {\bibfnamefont {S.}~\bibnamefont {Biscoveanu}}, \bibinfo
  {author} {\bibfnamefont {V.}~\bibnamefont {D’Emilio}}, \bibinfo {author}
  {\bibfnamefont {G.}~\bibnamefont {Ashton}},  \emph {et~al.},\ }\href
  {\doibase 10.1093/mnras/staa2850} {\bibfield  {journal} {\bibinfo  {journal}
  {Monthly Notices of the Royal Astronomical Society}\ } (\bibinfo {year}
  {2020}),\ 10.1093/mnras/staa2850},\ \bibinfo {note} {staa2850},\ \Eprint
  {http://arxiv.org/abs/https://academic.oup.com/mnras/advance-article-pdf/doi/10.1093/mnras/staa2850/33777524/staa2850.pdf}
  {https://academic.oup.com/mnras/advance-article-pdf/doi/10.1093/mnras/staa2850/33777524/staa2850.pdf}
  \BibitemShut {NoStop}%
\bibitem [{\citenamefont {{B\'ecsy, Bence and Raffai, Peter and Cornish, Neil
  J. and Essick, Reed and Kanner, Jonah and others}}(2017)}]{Becsy:2016ofp}%
  \BibitemOpen
  \bibfield  {author} {\bibinfo {author} {\bibnamefont {{B\'ecsy, Bence and
  Raffai, Peter and Cornish, Neil J. and Essick, Reed and Kanner, Jonah and
  others}}},\ }\href {\doibase 10.3847/1538-4357/aa63ef} {\bibfield  {journal}
  {\bibinfo  {journal} {Astrophys. J.}\ }\textbf {\bibinfo {volume} {839}},\
  \bibinfo {pages} {15} (\bibinfo {year} {2017})},\ \Eprint
  {http://arxiv.org/abs/1612.02003} {arXiv:1612.02003 [astro-ph.HE]}
  \BibitemShut {NoStop}%
\bibitem [{\citenamefont {Cotesta}\ \emph {et~al.}(2020)\citenamefont
  {Cotesta}, \citenamefont {Marsat},\ and\ \citenamefont
  {P\"urrer}}]{Cotesta:2020qhw}%
  \BibitemOpen
  \bibfield  {author} {\bibinfo {author} {\bibfnamefont {R.}~\bibnamefont
  {Cotesta}}, \bibinfo {author} {\bibfnamefont {S.}~\bibnamefont {Marsat}}, \
  and\ \bibinfo {author} {\bibfnamefont {M.}~\bibnamefont {P\"urrer}},\ }\href
  {\doibase 10.1103/PhysRevD.101.124040} {\bibfield  {journal} {\bibinfo
  {journal} {Phys. Rev. D}\ }\textbf {\bibinfo {volume} {101}},\ \bibinfo
  {pages} {124040} (\bibinfo {year} {2020})},\ \Eprint
  {http://arxiv.org/abs/2003.12079} {arXiv:2003.12079 [gr-qc]} \BibitemShut
  {NoStop}%
\bibitem [{\citenamefont {Hoy}\ and\ \citenamefont
  {Raymond}(2020)}]{Hoy:2020vys}%
  \BibitemOpen
  \bibfield  {author} {\bibinfo {author} {\bibfnamefont {C.}~\bibnamefont
  {Hoy}}\ and\ \bibinfo {author} {\bibfnamefont {V.}~\bibnamefont {Raymond}},\
  }\href@noop {} {\enquote {\bibinfo {title} {Pesummary: the code agnostic
  parameter estimation summary page builder},}\ } (\bibinfo {year} {2020}),\
  \Eprint {http://arxiv.org/abs/2006.06639} {arXiv:2006.06639 [astro-ph.IM]}
  \BibitemShut {NoStop}%
\end{thebibliography}%

\end{document}